\documentclass{pasa}

\usepackage{placeins}

\usepackage{graphicx}
\usepackage{amsmath, bm}   

\usepackage{chngcntr}  

\usepackage{booktabs}  

\graphicspath{{images/}{images/appendix/}{images/dense_gas/}{images/introduction/}{images/aa_plots/}}

\usepackage{xcolor}

\usepackage{xspace}

\newcommand{\kms}{\,km\,s$^{-1}$\xspace}
\newcommand{\vlsr}{v$_{\mathrm{lsr}}$\xspace}
\newcommand{\diff}{$\rm{cm}^2$\,s$^{-1}$\xspace}
\newcommand{\ergs}{$\rm{erg\,s^{-1}}$\xspace}

\newcommand{\hessj}{HESS\,J1804$-$216\xspace}
\newcommand{\snrg}{SNR\,G8.7$-$0.1\xspace}
\newcommand{\snr}{SNR\,8.3$-$0.1\xspace}
\newcommand{\psr}{PSR\,J1803$-$2137\xspace}
\newcommand{\psrn}{PSR\,J1803$-$2149\xspace}
\newcommand{\psrs}{PSR\,J1806$-$2125\xspace}
\newcommand{\fges}{FGES\,J1804.8$-$2144\xspace}
\newcommand{\OH}{1720\,MHz\,OH maser\xspace}

\title[\hessj]{Arc-minute-scale studies of the interstellar gas towards \hessj: Still an unidentified TeV $\gamma$-ray source}

\author[K. Feijen et al.]{K. Feijen$^1$\thanks{Email: kirsty.feijen@adelaide.edu.au}, G. Rowell$^1$, S. Einecke$^1$,
C. Braiding$^2$,
M. G. Burton$^{2,3}$,
N. Maxted$^{4,5}$,
F. Voisin$^1$,
G. F. Wong$^{6,7,5}$
\\
\affil{$^1$School of Physical Sciences, University of Adelaide, Adelaide, SA 5005, Australia}
\affil{$^2$School of Physics, University of New South Wales, Sydney, NSW 2052, Australia}
\affil{$^3$Armagh Observatory and Planetarium, College Hill, Armagh BT61 9DG, UK}
\affil{$^4$School of Science, The University of New South Wales, Australian Defence Force Academy, Canberra, 2600, Australia}
\affil{$^5$Western Sydney University, Locked Bag 1797, Penrith South DC, NSW 2751, Australia}
\affil{$^6$Pawsey Supercomputing Centre, 26 Dick Perry Ave, Kensington 6151, WA, Australia}
\affil{$^7$School of Physics, The University of New South Wales, Sydney 2052, Australia}
}

\jid{PASA}
\doi{10.1017/pas.\the\year.xxx}
\jyear{\the\year}

\usepackage{aas_macros}
\setcitestyle{aysep={}}
\usepackage{hyperref} 
\hypersetup{colorlinks,citecolor=blue,linkcolor=blue,urlcolor=blue}

\begin{document}

\begin{frontmatter}
\maketitle

\begin{abstract}
The Galactic TeV $\gamma$-ray source \hessj is currently an unidentified source. In an attempt to unveil its origin, we present here the most detailed study of interstellar gas using data from the Mopra Southern Galactic Plane CO Survey, 7 and 12\,mm wavelength Mopra surveys and Southern Galactic Plane Survey of HI. 
Several components of atomic and molecular gas are found to overlap \hessj at various velocities along the line of sight. The CS(1-0) emission clumps confirm the presence of dense gas. 
Both correlation and anti-correlation between the gas and TeV $\gamma$-ray emission have been identified in various gas tracers, enabling several origin scenarios for the TeV $\gamma$-ray emission from \hessj. 
For a hadronic scenario, \snrg and the progenitor SNR of \psr require cosmic ray (CR) enhancement factors of $\mathord{\sim} 50$ times the solar neighbour CR flux value to produce the TeV $\gamma$-ray emission. Assuming an isotropic diffusion model, CRs from both these SNRs require a slow diffusion coefficient, as found for other TeV SNRs associated with adjacent ISM gas.
The morphology of gas located at 3.8\,kpc (the dispersion measure distance to \psr) tends to anti-correlate with features of the TeV emission from \hessj, making the leptonic scenario possible. 
Both pure hadronic and pure leptonic scenarios thus remain plausible.

\end{abstract}

\end{frontmatter}


\section{INTRODUCTION}
\label{sec:intro}
\hessj is one of the brightest unidentified $\gamma$-ray sources, discovered by the High Energy Stereoscopic System (H.E.S.S.) in 2004 as part of the first H.E.S.S. Galactic Plane Survey \citep{VHE_Aharonian_2005}. \hessj features extended emission with a radius of $\mathord{\sim}22'$, a photon flux of almost 25\% of the Crab Nebula above 200\,GeV \citep{HESS_Aharonian_2006}, a TeV luminosity of $5\times10^{33}(d/\rm kpc)^2$\,\ergs and is one of the softest galactic sources with a photon index of $\Gamma = 2.69\,{\pm}\,0.04$ \citep{HGPS_2018}.

HAWC detected emission at $\mathord{\sim}4\sigma$ towards the north of \hessj, however no source has been identified.

\begin{figure}[!h]
\includegraphics[width=0.5\textwidth]{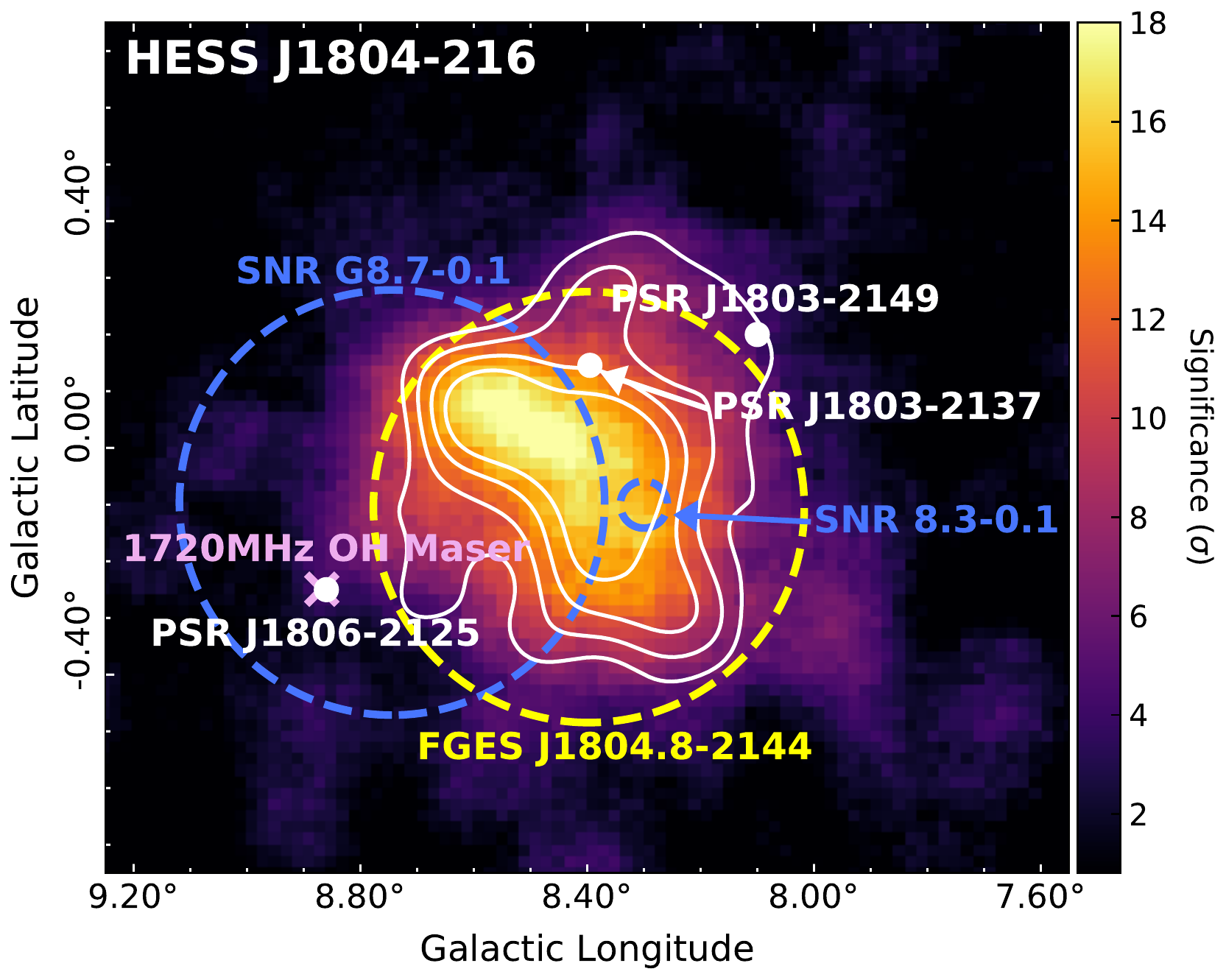}
\caption{TeV $\gamma$-ray significance map of \hessj, along with potential counterparts. \snrg and \snr are indicated by the blue dashed circles, \psr, \psrn and \psrs are indicated by the white dots and the \OH is indicated by a purple cross. \fges is shown by the yellow dashed circle. The TeV $\gamma$-ray emission for 5-10$\sigma$ is shown by the solid white contours. Image adapted from \cite{HGPS_2018}.}
\label{fig:excessCounts}
\end{figure}

The GeV $\gamma$-ray source, \fges, \citep{Fermi_Ackermann_2017} is a disk of radius $\mathord{\sim}23'$, coincident with the TeV emission from \hessj (see Figure~\ref{fig:excessCounts}).

\hessj has several possible counterparts found within $\mathord{\sim}1^{\circ}$ of its centroid, but none of these have been unambiguously associated with the TeV source. Two prominent candidates for the acceleration of cosmic rays (CRs) are supernova remnants (SNRs) and pulsar wind nebulae (PWNe). Here, the potential counterparts are: \snrg, \snr \citep[also referred to as SNR\,G8.3$-$0.0 in other literature, see][]{SNRs_Hewitt_2009}, \psr, \psrn and \psrs. The location of each counterpart with respect to \hessj is shown in Figure~\ref{fig:excessCounts}. The $\gamma$-ray contours used here were obtained from \cite{HESS_Aharonian_2006}.

\snr has radio shell-like morphology with a radius of $0.04^{\circ}$ \citep{CO_Kilpatrick_2016, Fermi1_Acero_2016}. \cite{CO_Kilpatrick_2016} find a systematic velocity of $+2.6$\,\kms, placing it at a kinematic distance of 16.4\,kpc, hence it is in the background. \snr would have an unusually high TeV luminosity \citep{HGPS_2018} at 16.4\,kpc of $1.34\times10^{36}$\,erg\,s$^{-1}$, making it unlikely to be powering \hessj.

\snrg has a large radius of $26'$ as determined by radio observations \citep{Old_SNR_Fang_2008}. It has been associated with a number of young H\textsc{ii} regions, forming the W30 complex, a large star forming region with a $\mathord{\sim}1^{\circ}$ region of radio continuum emission \citep{W30_Kassim_1990}. \snrg is a mature SNR with an age of 15\,kyr \citep{SNRage_Odegard_1986}. A distance of 4.5\,kpc is adopted here, which is found through X-ray observations and the column density of neutral hydrogen \citep{SNRs_Hewitt_2009}. \cite{Fermi_SNRG_2012} modelled the GeV to TeV emission assuming CRs are accelerated by this SNR.

A \OH is located along the southern edge of \snrg \citep{SNRs_Hewitt_2009}. It is currently categorised as an SNR-type maser, as no compact radio source has been found within $5'$ and it is believed to originate in a post-shock environment \citep{SNR_MCs_Fernandez_2013}. It is located at a velocity (\vlsr) of 36\kms corresponding to a distance of $\mathord{\sim}4.55$\,kpc, similar to the distance to \snrg. The coexistence of molecular clouds with \snrg and the location of the OH maser suggest that the SNR is interacting with nearby molecular clouds \citep{SNRs_Hewitt_2009}.

The characteristics of the pulsars are summarised in Table~\ref{tab:pulsars}. \psr was found by high-frequency radio observations by \cite{PSRs_Clifton_1986}. A dispersion measure distance of 3.8\,kpc is used here \citep{Xrays_Kargaltsev_2007}. \textit{Chandra} detected a faint and small ($\mathord{\sim}7'' \times4''$) synchrotron nebula around \psr, with the inner PWN positioned perpendicular to the direction of proper motion of the pulsar \citep{Xrays_Kargaltsev_2007}.

\psr is located towards the north-eastern edge of \snrg, but their association is highly unlikely according to a proper motion study of the pulsar \citep{PSRJ_Brisken_2006}. This study showed that for the pulsar to be born at the centre of \snrg a transverse velocity of $\mathord{\sim}1700$\kms is required. Therefore, \psr was born outside the central region of \snrg (see Figure~\ref{fig:PSR_motion}). The pulsar is most likely moving towards this area, rather than away from it, ruling out their connection \citep{PSRJ_Brisken_2006}. 

\psrs is a $\gamma$-ray quiet radio pulsar discovered with the Parkes radio telescope \citep{PSRJ1806_Parkes_2002}, and is located at a distance of $\mathord{\sim}10$\,kpc. Comparing the inferred $\gamma$-ray luminosity at 10\,kpc to the spin-down power, we obtain a TeV $\gamma$-ray efficiency ($\eta_{\gamma}=L_{\gamma}/\dot{E}$) of more than 100\%, excluding it as a plausible counterpart.

\psrn is a radio-quiet $\gamma$-ray pulsar located at a distance of 1.3\,kpc \citep{PSRn_Pletsch_2012}. This distance is obtained by inverting the $\gamma$-ray luminosity equation \citep[see][]{PSRn_Dist_2010} and is discussed further in Section~\ref{subsec:hadronic}. 

\begin{table*}[h]
\begin{center}
\caption{Pulsar characteristics, including spin-period (P), period derivative ($\dot{P}$), characteristic age ($\tau_c$), spin-down power~($\dot{E}$), distance and TeV luminosity at that distance.}
\begin{tabular}{ccccccc} 
\hline 
\addlinespace[0.1cm] 
Pulsar Name  & P   & $\dot{P}$    & $\tau_c$ & $\dot{E}$                & Distance & TeV Luminosity \\
{}    & (ms) & ($10^{-14}$) & (kyr)  & ($10^{35}$\,erg\,s$^{-1}$) & (kpc)   & ($10^{34}$\,erg\,s$^{-1}$) \\
\hline
J1803$-$2137$^a$ & 133.6 & 13.41        & 15.8     & 22.2   & 3.8  &  7.2 \\
J1803$-$2149$^b$ & 106.3 & 1.95         & 86.4     & 6.41   & 1.3  &  0.8 \\
J1806$-$2125$^c$ & 481.8 & 11.73        & 65.0     & 0.41   & 10   &  50 \\
\hline
\end{tabular}
\\ \small \textit{$^a$} From \cite{PSRJ_Brisken_2006},
\textit{$^b$} From \cite{PSRn_Abdo_2013},
\textit{$^c$} From \cite{PSRJ1806_Parkes_2002}
\label{tab:pulsars}
\end{center}
\end{table*}

Multiple studies \citep{HESSJ_Higashi_2008,HESSJ_Karg_2007, HESSJ_Lin_2013} have found a lack of X-ray emission towards \hessj, particularly towards \snrg and \psr. As mentioned previously, there is a faint and small X-ray nebula towards \psr. 
No SNR shell has been detected within the field of view of the \textit{Chandra} imaging \citep{HESSJ_Karg_2007}. 
Investigation of this region by \textit{XMM-Newton} \citep{HESSJ_Lin_2013} showed that the detected X-ray sources (both extended and point-like) are unlikely to be associated with \hessj due to them being located far away from the TeV peak.

Our detailed arc-min-scale ISM study here follows on from earlier work by \cite{1804_deWilt_2017} who revealed dense clumpy gas using the ammonia inversion line tracer. By studying the distribution and density of the ISM towards \hessj on arc-min scales, we can investigate morphological differences between hadronic and leptonic scenarios for the $\gamma$-ray production. We will utilise data from the Mopra radio telescope and Southern Galactic Plane Survey (SGPS) in order to carry out such an investigation, and look at an isotropic CR diffusion model for further insight into the likelihood of a hadronic interpretation.

\section{ISM OBSERVATIONS}
\label{sec:data_obs}
In this work, we utilised the publicly available Southern Galactic Plane Survey\footnote{Data can be found at \url{https://www.atnf.csiro.au/research/HI/sgps/fits_files.html}} of atomic hydrogen (HI) and 3, 7 and 12\,mm (frequency ranges 76-117\,GHz, 30-50\,GHz and 16-27\,GHz, respectively) data taken with the Mopra radio telescope towards the \hessj region.

The Australia Telescope Compact Array (ATCA) and Parkes telescope together mapped the HI emission along the Galactic Plane to form the SGPS. The survey is for latitudes of $b=\pm$1.5$^{\circ}$ and longitudes covering $l=253^{\circ}$\,-\,$358^{\circ}$ (SGPS I) as well as $l=5^{\circ}$\,-\,$20^{\circ}$ \citep[SGPS II,][]{SGPS_Naomi_2005}.

Mopra is a single dish with a 22\,m diameter surface. The 3\,mm data was taken from the Mopra Southern Galactic Plane survey, which is designed to map the fourth quadrant in the CO isotopologues (e.g. \citealt{Mopra_Braiding_2018}\footnote{Published Mopra data can be found at \url{https://dataverse.harvard.edu/dataverse/harvard/}}). The Mopra spectrometer (MOPS) was used in wide-band mode at 8\,GHz in Fast-On-The-Fly (FOTF) mapping to detect the four isotopologue lines ($^{12}$CO, $^{13}$CO, C$^{17}$O and C$^{18}$O). FOTF mapping is conducted by scanning across 1 square degree segments. To reduce artefacts in the data, each segment contains a longitudinal and latitudinal scan. The target region covering \hessj is $b=\pm 0.5^{\circ}$ and $l=7.0$\,-\,$9.0^{\circ}$ for the two CO isotopologue lines of interest; $^{12}$CO and $^{13}$CO.

The 7\,mm studies towards \hessj were taken in 2011 and 2012. The 7\,mm coverage is for a $49 \times 52$ arcmin region centred on $l=8.45^{\circ}$ and $b=-0.07^{\circ}$. MOPS was used in `zoom' mode for these observations. This provides 16 different sub-bands each with 4096 channels and a bandwidth of 137.5\,MHz \citep{Mopra_Urquhart_2010}. Table~\ref{tab:7mm} lists the various spectral lines at 7\,mm. 

The 12\,mm receiver on the Mopra telescope was used to carry out the H$_2$O Southern Galactic Plane Survey \citep[HOPS]{NH3_Walsh_2011}. This survey also detected other molecules such as the different inversion transitions of ammonia (NH$_3$). HOPS utilised On-The-Fly (OTF) mode with the Mopra wide-bandwidth spectrometer. HOPS mapped the region surrounding \hessj; $b=\pm 0.5^{\circ}$ and $l=7.0$\,-\,$9.0^{\circ}$.

The Mopra 3, 7 and 12\,mm data must be corrected to account for the extended beam efficiency of Mopra before any data analysis can be performed. The main beam brightness temperature is obtained by dividing the antenna temperature by the extended beam efficiency ($\eta_{XB}$). At 3\,mm (115\,GHz), for the CO(1-0) lines ($^{12}$CO and $^{13}$CO), a value of $\eta_{XB} = 0.55$ \citep{ATNF_Ladd_2005} is used. Following \cite{Mopra_Urquhart_2010}, the 7\,mm data are corrected to account for the beam efficiency of each frequency from Table~\ref{tab:7mm}. At 12\,mm for the NH$_3$(1,1) (24\,GHz) line the main beam efficiency of $\eta_{\rm{mb}} = 0.6$ is used  \citep{NH3_Walsh_2011}.

The Mopra data was processed using the Australia Telescope National Facility (ATNF) analysis software, \textsc{livedata}, \textsc{gridzilla}, and \textsc{miriad}\footnote{\url{http://www.atnf.csiro.au/computing/software/}}. Custom \textsc{idl} scripts were written to add further corrections and adjustments to the data \citep[see][]{Mopra_Braiding_2018}. \textsc{livedata} was used first to calibrate each map by the given OFF position and apply a baseline subtraction to the spectra. Next \textsc{gridzilla} was used to regrid and combine the data from each scan to create three-dimensional cubes (one for each molecular line in Table~\ref{tab:7mm}) of Galactic longitude, Galactic latitude and velocity along the line of sight (\vlsr). The produced FITS file is processed with both \textsc{miriad} and \textsc{idl}.

\section{SPECTRAL LINE ANALYSIS}
\label{sec:spec_lines}
\textsc{idl} and \textsc{miriad} were used to create integrated intensity maps. Different parameters, such as the mass and density, are calculated using these integrated intensity maps for each line described within this section. These parameters are examined to calculate important characteristics of each gas component towards \hessj (as shown in Sections~\ref{subsec:hadronic}~and~\ref{subsec:leptonic}). 

The mass of each gas region can be calculated, assuming that the gas consists of mostly molecular hydrogen with other constituents of the gas being negligible. The mass relationship is then given by: 
\begin{equation}
M=2 m_H N_{H_2} A
\end{equation}

where $m_H$ is the mass of a hydrogen atom, $N_{H_2}$ is the mean column density as obtained from each region and $A$ is the cross-sectional area of the region. The number density of the gas, $n$, is estimated using the area, $A$, column density, $N_{H_2}$, and volume, $V$ of the gas region, $n=N_{H_2}\,A/V$. For simplicity we assume a spherical volume for the clouds.

\subsection{Carbon monoxide}
\label{subsec:CO}
The focus for the 3\,mm study is the J=1-0 transition of the $^{12}$CO and $^{13}$CO lines. $^{12}$CO(1-0) is the standard molecule used to trace diffuse H$_{2}$ gas, as it is abundant and has a critical density of $\mathord{\sim}10^3$\,cm$^{-3}$ \citep{COdens_Bolatto_2013}. The CO brightness temperature is converted to column density with the use of an X-factor according to Equation~\ref{eqn:COX-factor}. 

\begin{equation}
N_{\rm{H_2}} = W_{\rm{CO}} \ X_{\rm{CO}} \quad \rm{cm}^{-2}
\label{eqn:COX-factor}
\end{equation}

Here, $N_{\rm{H_2}}$ is the column density of H$_2$, $W_{\rm{CO}}$ is the integrated intensity of the J=1-0 transition of either $^{12}$CO or $^{13}$CO and $X_{\rm{CO}}$ is a scaling factor with values presented in Equation~\ref{eqn:CO_X}, from \cite{ISM_Dame_2001} and \cite{MC_Simon_2001} for $^{12}$CO and $^{13}$CO respectively.

\begin{equation}
\begin{split}
X_{^{12}\rm{CO}} &= 1.8 \times 10^{20} \,  \rm{cm}^{-2}\,(\rm{K\,km/s})^{-1} \\
X_{^{13}\rm{CO}} &= 4.92 \times 10^{20} \, \rm{cm}^{-2}\,(\rm{K\,km/s})^{-1}
\end{split}
\label{eqn:CO_X}
\end{equation}

Since the $^{13}$CO(1-0) line is generally optically thin, as $^{13}$CO is 50 times less abundant than $^{12}$CO \citep{Mopra_Burton_2013}, the $^{13}$CO(1-0) line tends to follow denser regions of gas. The $^{13}$CO data will provide indication of the dense molecular gas components towards \hessj.

\subsection{Atomic hydrogen}
\label{subsec:HI}
The atomic form of hydrogen is detected through the 21\,cm line. The column density corresponding to a specific region is calculated through the relationship $N_{\rm{HI}} = W_{\rm{HI}} \ X_{\rm{HI}}$. Here the X-factor is from \cite{HI_Dickey_1990} (assuming the line is optically thin), as given by Equation~\ref{eqn:HI_X}.

\begin{equation}
X_{\rm{HI}} = 1.823 \times 10^{18} \, \rm{cm}^{-2}\,(\rm{K\,km/s})^{-1}
\label{eqn:HI_X}
\end{equation}

\subsection{Dense gas tracers}
As $^{12}$CO is one of the most abundant molecules in the universe it quickly becomes optically thick towards dense gas clumps. Tracers of dense gas ($n>10^4$\,cm$^{-3}$) are required to understand the internal dynamics and physical conditions of dense cloud cores. The following paragraphs outline the properties of various molecules used to trace the dense molecular clouds. These have a higher critical density and typically a much lower abundance compared to $^{12}$CO.

\paragraph{Carbon monosulfide}
Carbon monosulfide (CS) is far less abundant \citep{CS_Penzias_1971} than the other molecules previously mentioned and has a much higher critical density, on the order $10^4$\,cm$^{-3}$. The average abundance ratio between CS and molecular hydrogen is taken from \cite{CS_Frerking_1980} for quiescent gas to be $\mathord{\sim}10^{-9}$. CS is known to be a good tracer of dense molecular gas, especially in cases where the CO is optically thick. The focus here is CS(J=1-0) which is observable with the Mopra 7\,mm receiver.

\paragraph{Silicon Monoxide}
Similar to CS, silicon monoxide (SiO) is a tracer of dense gas and detectable via observing with a 7\,mm receiver. The SiO molecule originates in the compressed gas behind a shock moving through the ISM \citep{SiO_Martin_1992}. Such a shock can by found in star formation regions and in SNRs as they interact with the ISM \citep{SiO_Gusdorf_2008}. SiO can be a useful signpost of disruption in molecular clouds, where the SiO abundance is higher. \cite{W28_Nicholas_2012} detected clumps of SiO(1-0) towards various TeV sources, including the W28\,SNR. W28 shows a cluster of 1720\,MHz\,OH masers around the SiO emission, providing evidence of disrupted molecular clouds.

\paragraph{Methanol}
Methanol (CH$_3$OH) emission is a marker for star formation outflows and is an abundant organic molecule in the ISM \citep{CH3OH_Qasim_2018}. The CH$_3$OH line is often seen as a Class I maser. The detection of CH$_3$OH can be indicative of young massive stars and hence star formation regions. CH$_3$OH has also been detected in SNR shocks, where the gas is heated behind the shock front \citep{CH3OH_Voronkov_2010, W28_Nicholas_2012}.

\paragraph{Cyanopolyyne}
A cyanopolyyne is a long-chain of carbon triple-bonds (HC$_{2n+1}$N) found in the ISM often representing the beginning stages of high-mass star formation. The cyanopolyyne used here is cyanoacetylene, HC$_3$N. HC$_3$N is typically detected in warm molecular clouds and hot cores. It is present in dense molecular clouds and can be associated with star formation and H\textsc{ii} regions \citep{HC3N_2013}.

\paragraph{Ammonia}
The inversion transition of the ammonia molecule is denoted NH$_3$(J,K), for different quantum numbers J and K. NH$_3$ traces the higher density ($n\mathord{\sim}10^4$\,cm$^{-3}$) gas which can be associated with young stars \citep{NH3_Ho_1983, NH3_Walsh_2011}. It is readily observed in dense molecular clouds and towards various H\textsc{ii} regions. One common transition is NH$_3$(1,1) detected at a line frequency of $\mathord{\sim}23.69$\,GHz \citep{NH3_Walsh_2011}. The spectra of this inversion transition contains the main emission line surrounded by four weaker satellite lines. A study by \cite{1804_deWilt_2017} detected NH$_3$(1,1) emission towards H$_2$O masers in the vicinity of \hessj.

\section{RESULTS}
\label{sec:results}
The distribution and morphology of interstellar gas along the line of sight of the TeV $\gamma$-ray source \hessj is investigated in depth within this section. Multiple line emissions are analysed to investigate the characteristics of each ISM gas component along the line of sight. In particular we are interested in any spatial correlation or anti-correlation between the gas and the TeV $\gamma$-ray emission, as mentioned in Section~\ref{sec:intro}.

\subsection{Interstellar gas towards \hessj}
\label{subsec:ISM_1804}
\begin{figure}[!h]
\begin{center}
\includegraphics[width=\columnwidth]{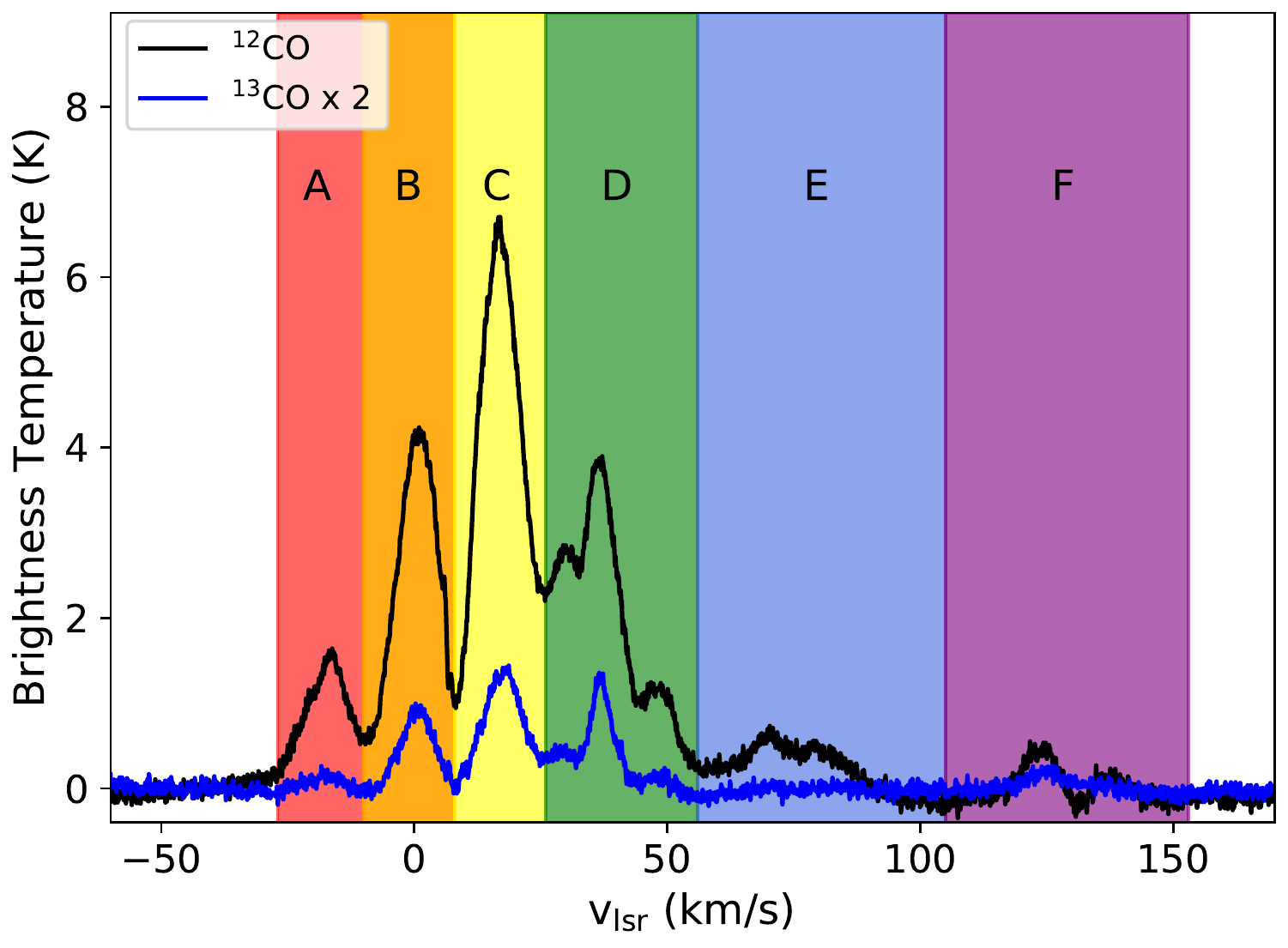}
\caption{CO(1-0) spectra towards \hessj with a radius of $0.42^{\circ}$ centred on $[l,b]=[8.4,-0.02]$ (see Figure \ref{fig:HII_reg}). Solid black lines and blue lines represent the emission spectra for Mopra $^{12}$CO(1-0) and $^{13}$CO(1-0) (scaled by a factor of 2) respectively. Velocity integration intervals for components A through F are shown by the coloured rectangles.}
\label{fig:velSpec}
\end{center}
\end{figure}

A circular region with a radius of $0.42^{\circ}$ which encompasses the extent of the TeV $\gamma$-ray emission from \hessj (shown by the cyan circle in Figure~\ref{fig:HII_reg}) is used to obtain spectra of the various molecular lines. The emission spectrum of the Mopra CO(1-0) data (Figure~\ref{fig:velSpec}) shows large regions of gas which overlap with \hessj and encompasses the bulk of its emission. Figure~\ref{fig:PVplot} shows a position-velocity (PV) plot of the Mopra $^{12}$CO(1-0) data, revealing the structure of the gas in velocity-space.

The CO(1-0) spectra show a large portion of the emission corresponds to a velocity range of \vlsr$\approx-40$\kms\,to\,$160$\,\kms. There are six main regions of emission along the line-of-sight as denoted by Table~\ref{tab:comp_v} and Figure~\ref{fig:velSpec}. The galactic rotation curve (GRC) model for \hessj (Figure~\ref{fig:1804GalRot}) is used to obtain `near' and `far' distances, based on the kinematic velocities to different ISM features.

\begin{table}[!h]
\centering
\caption{Velocity (\vlsr) integration intervals, with the corresponding distance measures, towards \hessj based on the components derived from the CO(1-0) spectra in Figure~\ref{fig:velSpec}.}
\setlength\tabcolsep{3.5pt}
\begin{tabular}{cccc} 
\hline 
Component & \vlsr  & Near Distance & Far Distance \\
          & (\kms) &    (kpc)      &  (kpc)       \\ 
\hline
A      & -27 to -10  &   0.1    &   ~17     \\
B      & -10 to 8    &   0.2    &   16.7    \\
C      &  8 to 26    &   3.0    &   13.8    \\
D      & 26 to 56    &   4.9    &   11.9    \\
E      & 56 to 105   &   6.4    &   10.4    \\
F      & 105 to 153  &   7.4    &   9.5     \\
\hline
\end{tabular}
\label{tab:comp_v}
\end{table}

The spectra for the HI data towards \hessj exhibits emission and absorption as shown in Figure~\ref{fig:velSpecHI}. Given HI is extremely abundant in the ISM, the data analysis will use the same velocity components as defined above from the CO data (Figure~\ref{fig:velSpec}).

\subsection{Discussion of ISM components}
\label{subsec:ISM_comps}
\begin{figure*}
\begin{center}
\includegraphics[width=1.6\columnwidth]{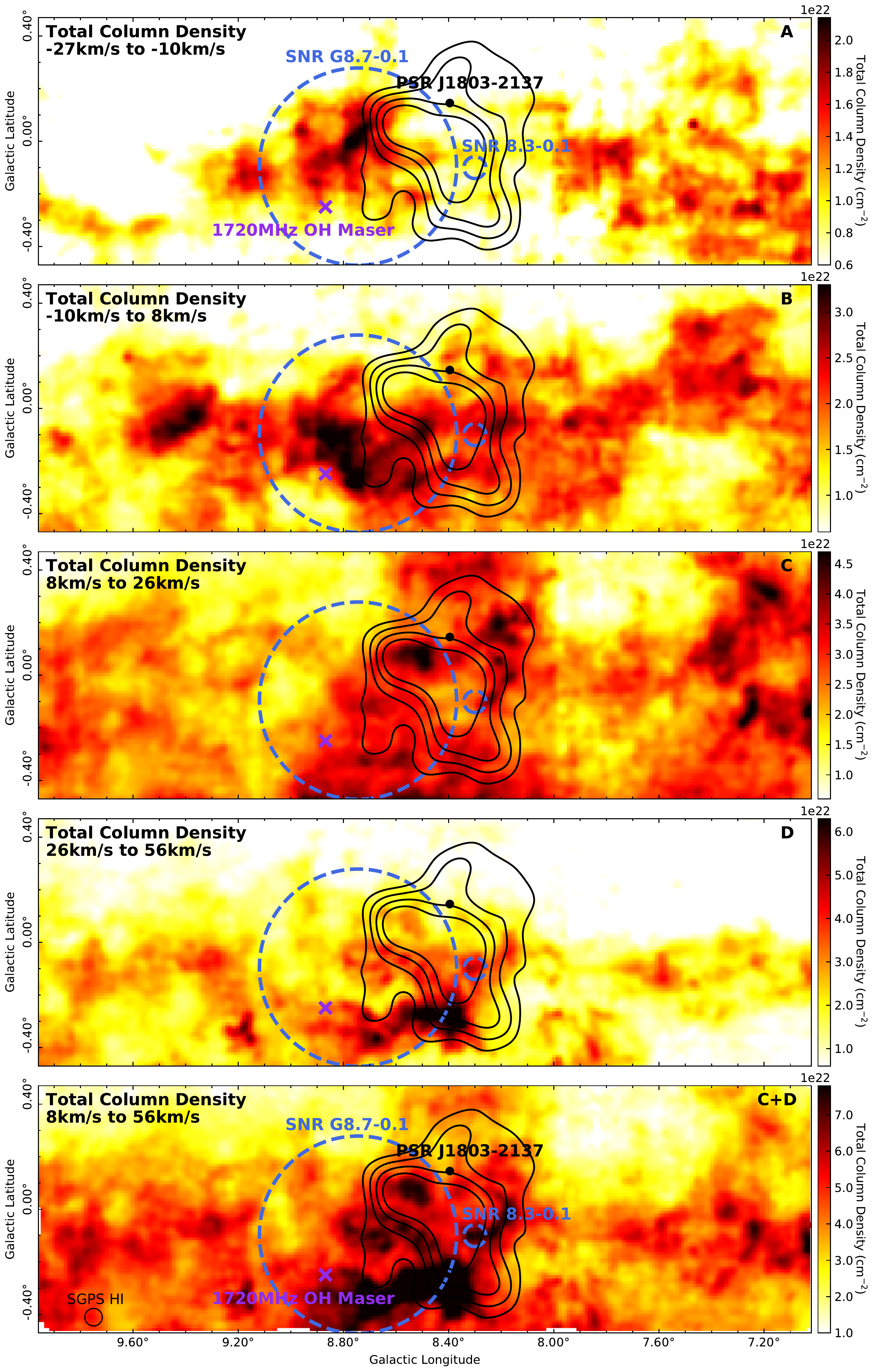}
\caption{Total column density maps, $2N_{\rm{H_2}}+N_{\rm{HI}}$, ($\rm{cm}^{-2}$) towards \hessj, for gas components A, B, C, D and C+D. The two dashed blue circles indicate \snrg and \snr. The \OH is indicated by the purple cross and \psr is indicated by the black dot. The TeV $\gamma$-ray emission for 5-10$\sigma$ is shown by the solid black contours.}
\label{fig:total}
\end{center}
\end{figure*}

It is important to look at both atomic and molecular hydrogen as they provide a look at the total target material available for CRs. The column density of both $^{12}$CO and HI are calculated using the X-factors from Equations~\ref{eqn:COX-factor}~and~\ref{eqn:HI_X} respectively. Maps of total column density for the selected integrated velocity ranges are essential in comparing the $\gamma$-ray emission and column density for the hadronic scenario. The total hydrogen column density, $N_{\rm{H}}$, is the sum of 2$N_{\rm{H_2}}$ and $N_{\rm{HI}}$ from Mopra $^{12}$CO (smoothed up to the beam size of the SGPS HI data) and SGPS HI observations respectively, giving the total proton content for each gas component. Figure~\ref{fig:ratio} shows the ratio between the column densities of molecular hydrogen and atomic hydrogen. This figure shows that the molecular gas tends to dominate over the atomic gas. The total column density maps for the defined velocity components are shown in Figure~\ref{fig:total}. This excludes components E and F (shown in Figure~\ref{fig:total_E+F}) as these have the weakest emission features and are distant. 

Figure~\ref{fig:total} also shows an extra component which covers the velocity range \vlsr$=8$\ to\ $56$\,\kms encompassing both components C and D, showing features that overlap much of \hessj. The dense gas structures in components C and D are connected by a lane of gas as shown in the position-velocity plot (Figure~\ref{fig:PVplot}). This indicates that some of the gas in these components are physically close to one another. The distances obtained from the galactic rotation model remain uncertain closer to the Galactic centre. Due to this, it is possible that the velocity/distance differences in component C and D (see Table~\ref{tab:comp_v}) arise from local motion. 

Figures~\ref{fig:12CO_mop},~\ref{fig:13CO_mop}~and~\ref{fig:HI_sgps} show mosaics of the integrated intensity maps of the Mopra $^{12}$CO(1-0), $^{13}$CO(1-0) and SGPS HI data respectively. The integrated intensity maps for the dense gas tracers are shown in the Appendix by Figures~\ref{fig:CS},~\ref{fig:SiO},~\ref{fig:HC3N},~\ref{fig:CH3OH}~and~\ref{fig:NH3}. The CS(1-0) and NH$_3$(1,1) will be discussed here. A number of H\textsc{ii} regions seen towards \hessj (see Figure~\ref{fig:HII_reg}) overlap with dense regions of interstellar gas, as discussed here.

\subsubsection{Component A}
\label{subsec:compA}
The $^{12}$CO(1-0) and $^{13}$CO(1-0) emission in component A (\vlsr$=-27$\ to\ $-10$\,\kms) show little overlap with \hessj. The emission in this component appears to be localised to the Galactic-West of the TeV source.

In HI there is a gas feature overlapping with the Galactic-East edge of \snrg which coincides with the central region of \hessj.

The NH$_3$(1,1) emission towards component A has no distinct features. 
The CS(1-0) data shows two dense features, one in the Galactic-North-East of \hessj and the other to the Galactic-South-East of the TeV source. The Galactic-North-East feature overlaps two H\textsc{ii} regions, G008.103$+$00.340 and G008.138$+$00.228, shown in Figure~\ref{fig:HII_reg}.

\subsubsection{Component B}
\label{subsec:compB}
In component B (\vlsr$=-10$\ to\ 8\,\kms) the $^{12}$CO(1-0) emission overlaps most of \hessj. There is gas filling the inner region of \snrg, with significant overlap with the Galactic-South-West to Galactic-West of the TeV source. This emission also extends West beyond both the SNR and TeV source. The $^{13}\rm{CO}(1\mbox{-}0)$ emission in this component follows a similar spatial morphology to the $^{12}$CO(1-0).

There is no HI overlap with \hessj for this component. The HI appears to anti-correlate with the $^{12}$CO(1-0) emission. 

There is an intense point-like region of NH$_3$(1,1) emission in the central region of \snrg, which corresponds to a maser detection in both CH$_3$OH and H$_2$O (see Figure~\ref{fig:NH3}). CS(1-0) emission in this component is quite weak.

\subsubsection{Component C}
\label{subsec:compC}
Component C (\vlsr=\ 8\ to\ 26\,\kms) shows some morphological matches between the $^{12}$CO(1-0) emission and the TeV $\gamma$-ray emission. There is however a depletion in molecular emission slightly south of the centre of \hessj (also seen in the $^{13}$CO(1-0) emission) which anti-correlates with the Southern TeV peak. Additionally, there is a prominent structure of gas running from Galactic-East to Galactic-West at the bottom of this panel (to the Galactic-South of the TeV source). Towards the Galactic-West of \hessj there is a molecular cloud which is positionally coincident with the northern edge of \snrg, as well as another clump of intense emission to the Galactic-East of this. Both of these features are also prominent in the $^{13}$CO(1-0) emission.
 
The HI emission (Figure~\ref{fig:HI_sgps}) appears to anti-correlate with the TeV $\gamma$-ray emission in component C, with very little emission detected in this area. Two clumps of HI gas overlap with the TeV source to the Galactic-North-West and East of \snr. In component C there is also a dense region of gas to the Galactic-North-West, the aforementioned clumps are not consistent with the $^{12}$CO(1-0) data.

The intense emission towards the Galactic-East of \psr in the total column density map (Figure~\ref{fig:total}) is also visible in both the CS(1-0) and NH$_3$(1,1) (Figures~\ref{fig:CS}~and~\ref{fig:NH3}). The significant CS(1-0) emission confirms the presence of dense gas in this region. This dense region is consistent with the infrared (IR) bright clouds and the H\textsc{ii} regions G008.103$+$00.340 and G008.138$+$00.228, as shown by Figure~\ref{fig:HII_reg}.

\subsubsection{Component D}
\label{subsec:compD}
In component D (\vlsr=\ 26\ to\ 56\,\kms), there is a distinct dense structure in the Galactic-South of \hessj present in both the $^{12}$CO(1-0) and $^{13}$CO(1-0) Mopra data. This dense emission overlaps with both \snrg and \hessj, so this region is likely to be associated with the SNR. This feature is consistent with several H\textsc{ii} regions; G008.362-00.303, G008.373-00.352, G008.438-00.331 and G008.666-00.351 (as indicated in Figures~\ref{fig:13CO_mop}~and~\ref{fig:HII_reg}). 

There is an intensity gradient in the CO emission as there is less gas towards the Galactic-North of this region. The CO emission toward the Galactic-North is weak and sparse. There is also weak emission seen outside \hessj towards the Galactic-West and Galactic-East. 

The HI emission shows a clear arm-like structure of emission that flows from the Galactic-East to Galactic-West through \hessj, most likely corresponding to the Norma Galactic Arm. This overlaps much of the central region of the source.

The NH$_3$(1,1) data for component D shows two distinct clumps in the Galactic-South which coincide with the previously discussed dense features from the molecular gas. These dense regions overlap with IR emission detected by the Spitzer GLIMPSE Survey in Figure~\ref{fig:HII_reg}. The IR emission is spatially coincident with several H\textsc{ii} regions. The clump outside \hessj is also traced by the CS(1-0) emission.

\subsubsection{Component E}
\label{subsec:compE}
In component E (\vlsr=\ 56\ to\ 105\,\kms) the $^{12}\rm{CO}(1\mbox{-}0)$ overlaps only a small portion of \hessj, corresponding to the central region of \snrg. There is a region of intense emission to the Galactic-North, near \psr. The $^{13}$CO(1-0) emission has less defined structure with no apparent overlap with the TeV source. 

The HI emission appears to have an arm like structure which extends from the Galactic-East to West of \hessj, with the denser regions towards the Galactic-West.

Both the NH$_3$(1,1) and CS(1-0) lines have almost no emission. A dense feature in the Galactic-South-West of \hessj overlaps the small H\textsc{ii} region G008.66$-$0.00351, shown in Figure~\ref{fig:HII_reg}.

\subsubsection{Component F}
\label{subsec:compF}
Both the $^{12}$CO(1-0) and $^{13}$CO(1-0) emission in component F (\vlsr=\ 105\ to\ 153\,\kms) show no overlap with the TeV source. This velocity component has little molecular emission aside from the clouds to the Galactic-South of \hessj.

A large HI feature overlaps \hessj, extending further to the Galactic-East in this component.

There is no significant NH$_3$(1,1) emission in component F. In the CS(1-0) data there is a dense core to the Galactic-South-East that has no spatial connection to the TeV $\gamma$-ray emission.

\section{DISCUSSION}
\label{sec:discussion}
Two different parent particle scenarios will be considered to be producing \hessj, a purely hadronic scenario and a purely leptonic scenario. As SNRs and PWNe are two candidates for accelerating CRs, the TeV $\gamma$-ray emission from \hessj could be the result of either scenario as both of these types are present within the field of view. The characteristics (i.e. mass and total column density) of the interstellar gas can be analysed to further investigate the complex nature of emission and to place a limit on which scenario is powering the TeV source.

\subsection{Purely Hadronic Scenario}
\label{subsec:hadronic}
The hadronic production of TeV $\gamma$-rays involves the interaction of CRs and matter in the ISM. A study by \cite{Old_SNR_Yamazaki_2006} showed that old SNRs tend to have a large enough hadronic contribution to account for the TeV $\gamma$-ray emission. This is seen both at the SNR shock location and at the associated molecular clouds. 

\paragraph{CRs from \snrg}
Many 1720\,MHz\,OH masers have been seen towards other TeV $\gamma$-ray SNRs, such as W28, W44 and IC\,443 \citep{W28_Frail_1994,MaserMC_Claussen_1997}, which provides evidence of interaction between the SNR shock and molecular clouds surrounding it \citep[e.g.][]{W28_Nicholas_2012}. The presence of the \OH towards \snrg is consistent with CRs being accelerated by this SNR. Therefore, this section will assume that \snrg is the accelerator of hadronic CRs.

To test whether a hadronic scenario is initially feasible, the total energy budget of CRs, $W_{p,\rm{TeV}}$, is calculated using:
 
\begin{equation}
W_{p,\rm{TeV}}=L_{\gamma} \tau_{pp}
\label{eqn:tot_E_budget}
\end{equation}

where $L_{\gamma}$  is the luminosity of the $\gamma$-ray source. The TeV $\gamma$-ray luminosity varies depending on the distance to each counterpart, $L_{\gamma}\mathord{\sim}5\times10^{33}(d/\rm kpc)^2 \,erg\,s^{-1}$.
The cooling time of proton-proton collisions is given by \cite{Pi0decay_Aharonian_1996}:

\begin{equation}
\tau_{pp} = 6 \times 10^7 \, ( n\rm{/cm}^{-3} )^{-1} \, \rm{yr}
\label{eqn:tau_pp}
\end{equation}
 
where $n$ is the number density of the target ambient gas, found in a circular region which encompasses the TeV $\gamma$-ray contours of \hessj above $5\sigma$, with a radius of $0.42^{\circ}$. 

Another relationship can be made between the amount of CRs that are incident upon the gas and the $\gamma$-ray flux $F(>E_{\gamma})$ above some energy $E_{\gamma}$. The CRs have diffused through the ISM allowing the spectra to steepen from an $E^{-2}$ power law at the accelerator to $E^{-2.6}$ at some distance from the CR source. Therefore, we assume an $E^{-1.6}$ integral power law spectrum from the integration of $dN_p=E^{-2.6}\,dE_p$, as given by \citep{VHE_Aharonian_1991}:

\begin{equation}
F(\geq E_{\gamma})=2.85 \times 10^{-13} E_{\rm{TeV}}^{-1.6} \left( \dfrac{M_5}{d_{\rm{kpc}}^2} \right) k_{\mathrm{CR}} \quad \rm{cm}^{-2}\,\rm{s}^{-1}
\label{eqn:CR_flux_k}
\end{equation}

The photon flux for $\gamma$-rays from \hessj is $F(\ge200\,\rm GeV)=5.32\times 10^{-11}\,cm^{-2}\,\rm{s}^{-1}$ \citep{HESS_Aharonian_2006}. The distance to the gas component in kpc is $d_{\rm{kpc}}$ and $M_5$ is the mass of the CR target material in units $10^5$ M$_{\odot}$. The CR enhancement factor, $k_{\rm CR}$, is the ratio of the CR flux at the ISM interaction point compared to that of Earth-like CR flux.

The maps of total column density ($2N_{\rm{H_2}}+N_{\rm{HI}}$) in Figure~\ref{fig:total} were used to find the mean column density of each velocity component in order to calculate both the number densities and masses of each velocity component. Equations~\ref{eqn:tot_E_budget}~and~\ref{eqn:CR_flux_k} are used to calculate the total CR energy budget ($W_{p,\rm{TeV}}$) and the CR enhancement factor ($k_{\rm CR}$) for each gas component respectively (shown in Table~\ref{tab:tot_CR_budget}).

\begin{table*}[h]
\begin{center}
\caption{CR enhancement values, $k_{\rm CR}$ (Equation~\ref{eqn:CR_flux_k}), and total energy budget of CRs, $W_{p,\rm{TeV}}$ (Equation~\ref{eqn:tot_E_budget}), for each velocity component defined in Figure~\ref{fig:velSpec}. Each of these numbers are calculated from the maximum extent of \hessj (circle of radius $0.42^{\circ}$). The values for total mass and and column density are obtained from the total column density of hydrogen, using the $^{12}$CO and HI data from Mopra and SGPS respectively.  The near distances were derived using the GRC presented in Figure~\ref{fig:1804GalRot}. The magnetic field is calculated using Equation~\ref{eqn:B_crutcher}.}
\begin{tabular}{ccccccccc}
\hline
Component & $d$ &  $n$        & $N_{\rm H}$            & $M$       & $k_{\rm CR}$ & $W_{p,\rm{TeV}}$     & $B$ \\
       & (kpc)  & (cm$^{-3}$) & (10$^{21}$\,cm$^{-2}$) & $(10^4$\,M$_{\odot})$ &  & (10$^{48}$\,erg)     & ($\mu$G) \\
\hline
A     &   0.1  &  2835  &  8.6   &  0.01  &  123  &  $3\times10^{-5}$  &  43 \\
B     &   0.2  &  4385  &  18.8  &  0.05  &  56   &  $4\times10^{-5}$  &  57 \\
C$^a$ &   3.8  &  325   &  29.4  &  36    &  57   &  0.5      &  11 \\
D$^b$ &   4.5  &  160   &  23.7  &  79    &  37   &  1.1      &  10 \\
C+D   &   4.4  &  400   &  52.9  &  138   &  20   &  0.5      &  12 \\
E     &   6.4  &  25    &  4.8   &  27    &  221  &  15.8     &  10 \\
F     &   7.4  &  5     &  1.3   &  9     &  842  &  91.8     &  10 \\
\hline
\end{tabular}
\\ \small The values of distance are taken from the kinematic velocity average of each component. 
\\ \small \textit{$^a$} Component C values are taken specifically for \psr.
\\ \small \textit{$^b$} Component D values are taken specifically for \snrg.
\label{tab:tot_CR_budget}
\end{center}
\end{table*}

An SNR has a total canonical kinetic energy budget of $\mathord{\sim}10^{51}$\,erg, of this we expect an amount of $\mathord{\sim}10^{50}\,\rm{erg}$ ($\mathord{\sim}10\%$) to be converted into CRs. From Table~\ref{tab:tot_CR_budget} the total energy budget for components C, D and C+D are on the order of $W_{p,\rm{TeV}}=10^{48}$\,erg which suits the criteria of being $<10^{50}$\,erg. The values of $k_{\rm CR}$ for these ISM components are on the order of $\mathord{\sim}10$, which is acceptable provided we have a young to middle aged (10$^3$\ to\ $10^5$\,yrs) impulsive CR accelerator within 10\ to\ 30\,pc of the target material \citep{Pi0decay_Aharonian_1996}. 

At a distance of 4.5\,kpc, \snrg is placed at a kinematic velocity of $\mathord{\sim}$35\,\kms according to the GRC (outlined in Appendix \ref{subsec:GRC}). This corresponds to component D as shown in Table~\ref{tab:comp_v}. The values for total energy budget, $W_{p,\rm{TeV}}$, in Table~\ref{tab:tot_CR_budget} are considered as a lower limit on the total CR energy budget as we are considering $\gamma$-rays of energies above 200\,GeV corresponding to CR energies of $\mathord{\sim}$1.2\,TeV \citep[from the relation $E_{\gamma}\mathord{\sim}0.17E_{\rm{CR}}$, in][]{X_Kelner_2006}. For \snrg (component D) we require a CR enhancement factor, $k_{\rm CR}$, of $\mathord{\sim}37$ times that of the Earth-like CR density to produce the observed $\gamma$-ray flux towards \hessj for a hadronic scenario to be plausible.

The total column density map (Figure~\ref{fig:total}) shows the ISM partially overlapping the TeV $\gamma$-ray emission from \hessj. This cloud shows a good morphological match with component D (see Figure~\ref{fig:total}), corresponding to the distance of \snrg. It is therefore possible that this cloud is a target for CRs generated by \snrg.

\vspace{4mm}
Following \cite{Pi0decay_Aharonian_1996}, the volume distribution of CRs ($\rm{cm^{-3}\,GeV^{-1}}$) as a function of the injection spectrum, $N_0 E^{-\alpha}$, is given by Equation~\ref{eqn:inj_spec}. This assumes a spherically symmetric case for the diffusion equation, in which relativistic particles accelerated by a source, escape and enter the ISM.

\begin{equation}
f(E,R,t) \approx 
\dfrac{N_0 E^{-\alpha}}{\pi^{3/2}R_{\rm{dif}}^3}
\exp \left( -\dfrac{(\alpha-1)t}{\tau_{pp}} - \dfrac{R^2}{R_{\rm{dif}}^2} \right)
\label{eqn:inj_spec}
\end{equation}

where the diffusion radius

\begin{equation}
R_{\rm{dif}} \equiv R_{\rm{dif}}(E,t) = 
2 \sqrt{D(E)t \dfrac{\exp(t\delta/ \tau_{pp})-1}{t\delta/\tau_{pp}}}
\label{eqn:R_dif}
\end{equation}
 
is the radius given for CR protons of energy $E$ propagating though the ISM during time $t$. The proton-proton cooling time, $\tau_{pp}$, is given by Equation~\ref{eqn:tau_pp} and $\alpha=2$. We consider a specific CR accelerator model in which the age of \snrg is taken to be 15\,kyr and 28\,kyr from \cite{SNRage_Finley_1994}. The diffusion coefficient, $D(E)$, is determined using Equation~\ref{eqn:diff_coefficient} from \cite{Lep_Gabici_2007}.

\begin{equation}
D(E) = \chi D_0 \left( \dfrac{E/\rm{GeV}}{B/3\mu \rm{G}}  \right)^{\delta}
\label{eqn:diff_coefficient}
\end{equation}

where $\chi$ is a diffusion suppression factor (typically $\chi$<1 inside a molecular cloud). The factor $\chi$ from \cite{Pi0decay_Aharonian_1996} takes values of 0.01 and 1 to represent `slow' and `fast' diffusion respectively. A value of $\chi=$0.01 is usually taken to account for the dense regions of interstellar gas that CRs may diffuse through. 
Various diffusion suppression factors have been found through different studies on the W28\,SNR \citep{Chi_Hui_2010, Chi_Giuliani_2010, chi_Gabici_2010}. \cite{Chi_Hui_2010} assume $\chi=0.1$, \cite{Chi_Giuliani_2010} use $\chi=0.01$, while \cite{chi_Gabici_2010} adopt a value of $\chi=0.06$. It is clear that the diffusion suppression factor is poorly constrained. Here, we adopt a value from $\chi=0.001$\ to\ 0.1.
The index of diffusion coefficient, $\delta$ is typically given a value of $0.3$\,-\,$0.7$ \citep{Book_Berezinskii_1990}.
$D_0$ and $\delta$ are given the galactic values of $3\times 10^{27}\,\rm{cm}^2\,\rm{s}^{-1}$ and $0.5$, respectively, whilst 3\,$\mu \rm{G}$ is the Galactic disc's average magnetic field.
\cite{Mag_Crutcher_2010} gives a relationship between the magnetic field $B$ and number density $n$ of a given region, shown by Equation~\ref{eqn:B_crutcher}. They found that the magnetic field is enhanced in dense ($n>300\,\rm{cm}^{-3}$) molecular clouds. 

\begin{equation}
B =
\begin{cases}
B_0               \quad & \mathrm{for} \, n < 300\,\rm{cm}^{-3} \\
B_0(n/n_0)^{0.65} \quad & \mathrm{for} \, n > 300\,\rm{cm}^{-3}
\end{cases}
\label{eqn:B_crutcher}
\end{equation}

where $n$ is the number density in the cloud, $n_0$ is a constant number density set to 300\,cm$^{-3}$, $B$ is the maximum magnetic field in the cloud and $B_0\mathord{\sim}10 \mu \rm{G}$. The various magnetic field values for each ISM component are shown in Table~\ref{tab:tot_CR_budget}.

The normalisation factor, $N_0$, is determined assuming the SNR is at an early epoch of evolution ($\mathord{\sim}1$\,yr) meaning $R_{\rm{dif}}$ is approximated by the size of the SNR (i.e. $R_{\rm{dif}} = R$). 
The CR energy produced by the SNR is $<10^{50}$\,erg. It is taken here to be $2\times10^{48}$\,erg to match the observed GeV and TeV CR enhancement factors as shown in Figure~\ref{fig:aa_snr_in}. We note that $N_0$ is considered a lower limit since the $k_{\rm CR}$ constant from Equation~\ref{eqn:CR_flux_k} assumes all of the cloud mass is impacted by CRs and converted to $\gamma$-rays. Energy-dependent diffusion and penetration \citep[e.g.][]{Lep_Gabici_2007} inside the dense clouds, highlighted by the $^{13}$CO peaks in Figure~\ref{fig:13CO_mop}, could however infer a higher $k_{\rm CR}$ value.
In addition, clouds are typically not physically connected, given the typically wide range of distances inferred from the cloud velocities spanning Galactic arms (c.f. Figure~\ref{fig:PVplot}).

The initial power-law distribution is assumed to be $dN/dE=E^{-2}$ for determining $N_0$.

\begin{figure}[!h]
\begin{center}
\includegraphics[width=\columnwidth]{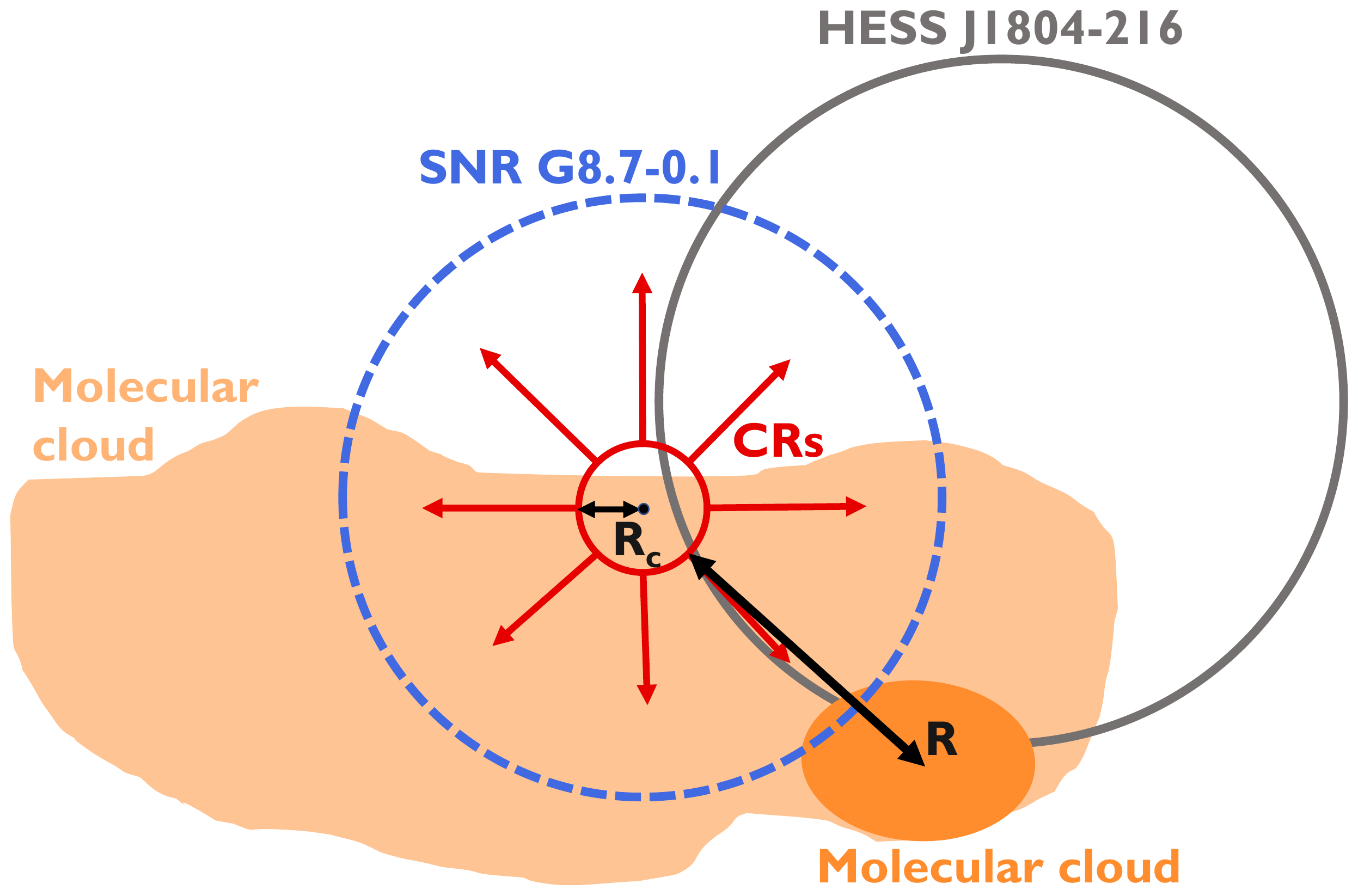}
\caption{Schematic of CRs escaping \snrg before interacting with the molecular clouds in component D to create the TeV $\gamma$-ray emission from \hessj. The red circle shows the release point of CR protons at a radius of $R_{\rm{c}}\mathord{\sim}5$\,pc. The black line shows the physical distance between the cloud and the release point of CRs ($R\mathord{\sim}12$\,pc).}
\label{fig:SNR_CR}
\end{center}
\end{figure}

\begin{figure}[!h]
\begin{center}
\includegraphics[width=\columnwidth]{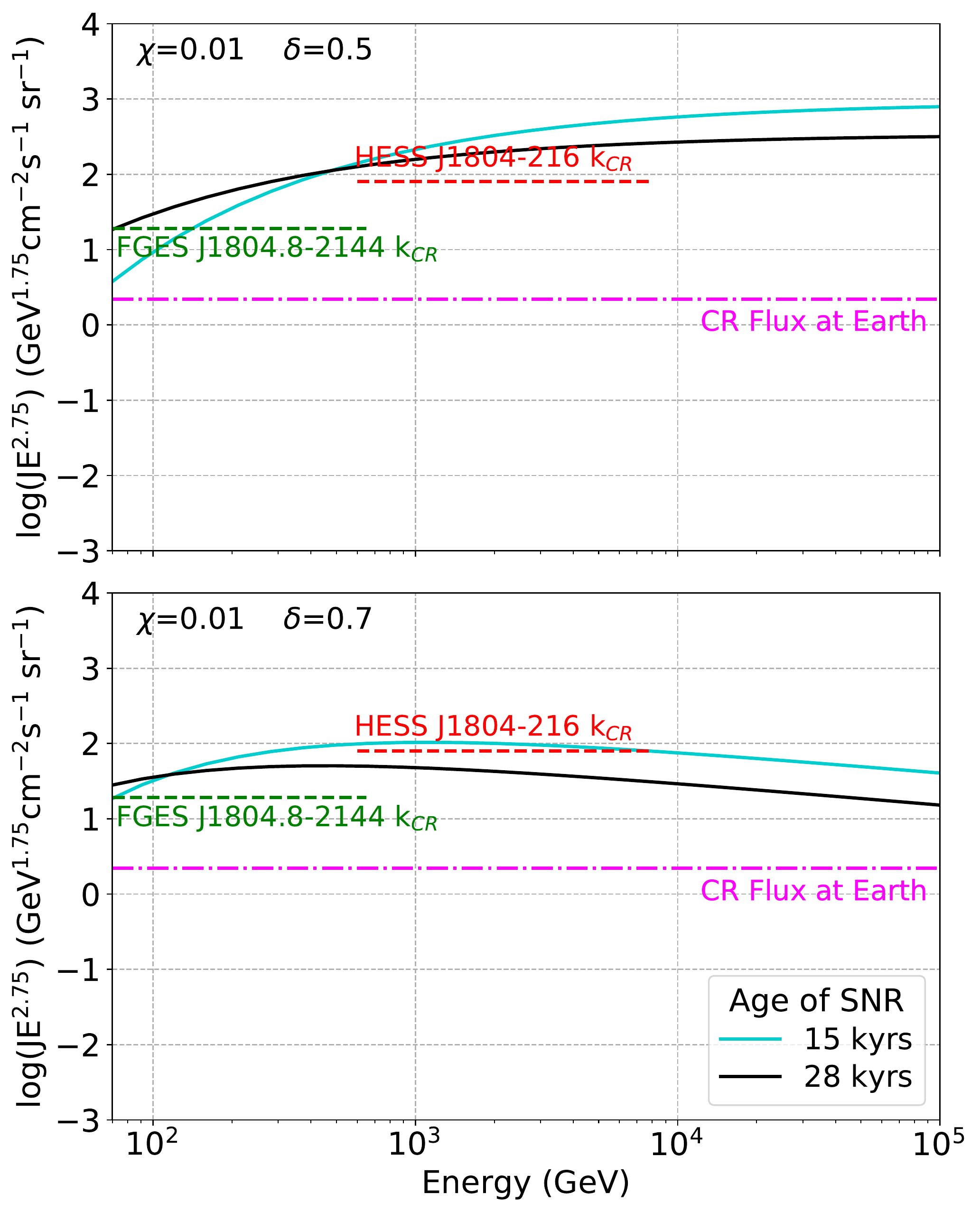}
\caption{Modelled energy spectra of CR protons (Equation~\ref{eqn:flux_dif}) escaping from a potential impulsive accelerator (e.g. \snrg), with a total energy of $2\times10^{48}$\,erg in CRs. The model shows different values for the diffusion suppression factor, $\chi$, and index of the diffusion coefficient, $\delta$.
A power-law spectrum with an index of $\alpha=2$ is assumed. The number density is taken to be $n=160$\,cm$^{-3}$. The distance from the accelerator to the cloud is $R\mathord{\sim}12$\,pc and age of the source are taken to be 15\,kyr and 28\,kyr for the cyan and black curves, respectively. The magenta dashed line represents the CR flux observed at Earth. The red dashed line represents the calculated CR enhancement factor for \hessj ($k_{\rm CR}\approx37$). The green dashed line  represents the calculated CR enhancement factor for \fges ($k_{\rm CR}\approx9$).}
\label{fig:aa_snr_in}
\end{center}
\end{figure}

The radius of the SNR shock during the Sedov phase (when mass of the swept-up material exceeds the mass of the supernova ejecta) is given by Equation~\ref{eqn:r_SNR} \citep{SNR_Reynolds_2008}. 
\begin{equation}
R_{\rm{c}} = 0.31 
\left( \dfrac{E_{51}}{n_0} \right) ^{1/5} 
\left( \dfrac{\mu_1}{1.4} \right) ^{-1/5} 
t_{\rm{yr}}^{2/5} \  \rm{pc}
\label{eqn:r_SNR}
\end{equation}

where $E_{51}$ is the ejected supernova kinetic energy in units of $10^{51}$\,erg, $n_0$ is the number density, $\mu_1$ is the mean mass per particle \citep[taken to be 1.4, from][]{SNR_Reynolds_2008} and $t_{\rm{yr}}$ is the escape time of CR protons. We assume $R_{\rm{c}}$ is the radius at which CR protons are released from the accelerator, which can then be used to calculate the distance to the cloud in component D. For \snrg we assume $E_{51}=1$, with a number density of $n_0=160$\,cm$^{-3}$ for component D. The escape time of CRs \citep[e.g.][]{Gabici_2009} from a SNR shock is:

\begin{equation}
t_{\rm{esc}} = t_{\rm{Sedov}}
\left(
\dfrac{E_\text{p}}{E_{\text{p,max}}}
\right) ^{-1/\delta_p}
\label{eqn:t_esc}
\end{equation}

where the maximum energy of CR protons is $E_{\rm{p,max}}$=500\,TeV, $t_{\rm{Sedov}}$=100\,years,  $E_{\rm{p}}$=150\,TeV and $\delta_p=0.5$ \citep{tesc_Casanova_2010}. 
Using Equation~\ref{eqn:r_SNR} the release point of the CRs is taken to be $R_{\rm{c}}\mathord{\sim}5$\,pc, therefore the physical distance to the cloud from this point is $R\mathord{\sim}12$\,pc. Figure~\ref{fig:SNR_CR} shows a schematic of this scenario where CRs are accelerated by \snrg and escape before interacting with the nearby cloud structure defined by component D (Figure~\ref{fig:total}). 

The differential flux of CR protons is then given by:

\begin{equation}
J(E,R,t) = (c/4\pi) f(E,R,t) \quad \rm{cm^{-2}}\,s^{-1}\,GeV^{-1}\,sr^{-1}
\label{eqn:flux_dif}
\end{equation}

Figure~\ref{fig:aa_snr_in} shows the derived energy spectrum of CR protons escaping from \snrg from Equation~\ref{eqn:inj_spec}.  
The scenario assumes that \snrg is an impulsive accelerator meaning the bulk of CRs escape the SNR at $t=0$, compared to the continuous case in which CRs are continuously injected in the ISM. The CR enhancement factors for component D are shown for TeV energies (from Equation~\ref{eqn:CR_flux_k} and Table~\ref{tab:tot_CR_budget}) and GeV energies ($k_{\rm CR}\mathord{\sim}9$ from Equation~\ref{eqn:CR_flux_k2}). Here we show the two cases that broadly match the observed GeV and TeV CR enhancement factors where $\delta$ is 0.5 or 0.7 (Equation~\ref{eqn:R_dif}) and $\chi=0.01$. The parameters $\delta$ and $\chi$ were varied, as shown in Appendix \ref{subsec:A_A_plots} (Figure~\ref{fig:aa_vary_delta}), until a reasonable match was found. The contribution from the spectrum of CR protons observed at Earth \cite[i.e. in the solar neighbourhood from][]{EarthCR_Dermer_1986}, as given by Equation~\ref{eqn:flux_earth}, is also shown.

\begin{equation}
J_{\odot}(E) = 2.2 E^{-2.75} \quad \rm{cm^{-2}}\,s^{-1}\,GeV^{-1}\,sr^{-1}
\label{eqn:flux_earth}
\end{equation}

The results in Figure~\ref{fig:aa_snr_in} show that the older age assumption for \snrg (28\,kyr) has a lower energy population of CRs, and the higher energy CRs are seen to escape first, as expected. 
The total CR energy budget of $2\times10^{48}$\,erg is consistent with $W_{p,\rm{TeV}}$ from Equation~\ref{eqn:tot_E_budget} (c.f. Table~\ref{tab:tot_CR_budget}) as computed using component D and tends to match the observed CR enhancement factors.
It is evident that the pure hadronic scenario requires slow diffusion ($\chi \le 0.01$) in order to contribute to the $\gamma$-ray emission for \hessj. Small values of $\chi$<0.05 are noted in other studies \citep{Chi_Hui_2012, tdiff_Protheroe_2008,chi_Gabici_2010} for various sources including the W28\,SNR, W44 and IC443, all with similar ages to \snrg. 
Our diffusion index of $\delta$ in the range 0.5 to 0.7 is consistent with \cite{Fermi_SNRG_2012} who found a diffusion index of $\delta=0.6$ from their modelling of the GeV to TeV emission. We note that the GeV emission position now overlaps the TeV position \citep{Fermi_Ackermann_2017}, whereas previously \citep[in][]{Fermi_SNRG_2012} the GeV emission was located closer to \snrg.

In Figure~\ref{fig:aa_snr_in} both ages tend to match the CR enhancement factors for \hessj and \fges.

\paragraph{CRs from the progenitor SNR of \psr}
\psr currently has no known SNR associated with it. Here we discuss the possibility that the undetected progenitor SNR from \psr is accelerating CRs. Using the hadronic scenario outlined above we assume the centre of this SNR is located at the birth position of \psr, placing it in gas component C (consistent with \psr). We assume the progenitor SNR is 16\,kyr old, consistent with the age of \psr. A distance of $\mathord{\sim10}$\,pc is used as the distance from the release point of CRs to the cloud to the Galactic-South-West of \psr in component C.
The model in Figure~\ref{fig:aa_psr} shows the energy spectrum of CR protons escaping from the progenitor SNR of \psr. The CR enhancement factors for component C are shown for TeV energies ($k_{\rm CR}\mathord{\sim}57$) and GeV energies ($k_{\rm CR}\mathord{\sim}14$ from Equation~\ref{eqn:CR_flux_k2}). A $\chi$ value of 0.01 for $\delta=0.5,\ 0.7$ or $\chi=0.001$ for $\delta=0.7$, could potentially match the observed values from \hessj and \fges.

\subsection{Purely Leptonic Scenario}
\label{subsec:leptonic}
\paragraph{\psr powered PWN}
Here we consider TeV $\gamma$-ray emission produced by high-energy (multi-TeV) electrons primarily interacting with soft photon fields via the inverse-Compton process. \psr is located $\mathord{\sim}0.2^{\circ}$ from the centre of \hessj (as seen in Figure~\ref{fig:excessCounts}). \psr is at a distance of 3.8\,kpc (see Section~\ref{sec:intro}) which corresponds to a velocity of $\mathord{\sim}25$\,\kms, placing this pulsar in gas component C. Due to the extended nature of \hessj, and the high spin down luminosity of \psr, it is possible that the TeV emission is produced by high energy electrons from \psr as a PWN. A recent study \citep{PWNe_Aharonian_2018} shows 14 firmly identified PWNe contribute to the TeV population of H.E.S.S. sources.

The spin down luminosity of \psr ($\dot{E}=2.2 \times 10 ^{36}\,\rm{erg}\,\rm{s}^{-1}$) is compared with the $\gamma$-ray luminosity of \hessj $L_{\gamma} = 7.1 \times 10 ^{34}\,\rm{erg}\,\rm{s}^{-1}$ at 3.8\,kpc, to obtain a TeV $\gamma$-ray efficiency of $\eta_{\gamma}=L_{\gamma}/\dot{E}\sim3\%$. 
This is consistent with the typical efficiency of pulsars (potentially) associated with TeV sources according to \cite{PWNe_Aharonian_2018}, meaning leptonic $\gamma$-ray emission from a PWN is supported from an energetics point of view.

In the scenario of a PWN-driven TeV $\gamma$-ray source, the TeV emission is expected to anti-correlate with the surrounding molecular gas. 
High energy electrons suffer significant synchrotron radiation losses due to the enhanced magnetic field strength in molecular clouds, leading to anti-correlation between the gas and $\gamma$-rays.
Assuming the gas in component C is located at the same distance as \psr, there is indeed some anti-correlation between the total column density in Figure~\ref{fig:total} and the TeV emission towards the Galactic-South of the TeV peak.

To account for the observed TeV $\gamma$-ray emission electrons must be able to diffuse across the extent of the GeV and TeV sources. Electrons are therefore required to travel a distance of $R\mathord{\sim}30$\,pc from \psr to the nearby cloud in component C (see Figure~\ref{fig:total}). The radiative cooling times are calculated based on the assumption that electrons are being accelerated by \psr. 

The inverse-Compton cooling time $t_{\rm{IC}}$ in the Thomson regime is given by:

\begin{equation}
t_{\rm{IC}} \approx 3 \times 10^8 (U_{\textrm{rad}}/\textrm{eV}/\textrm{cm}^3)^{-1} 
			(E_e /\textrm{GeV})^{-1}\, \textrm{yr}
\label{eqn:t_IC}
\end{equation}

where $U_{\rm{rad}}$ is $0.26\,\rm eV/cm^3$ (the energy density of the Cosmic Microwave Background). For any given H.E.S.S. source we expect 100\,GeV $\gamma$-rays (the lower limit detectable by H.E.S.S.) as produced by inverse-Compton scattering to correspond to electron of energies of $E_e\mathord{\sim}6$\,TeV ($E_e\mathord{\sim}20\sqrt{E_{\gamma}}$ for the Thomson scattering regime). 

The synchrotron cooling time $t_{\rm{sync}}$ is given by:

\begin{equation}
t_{\rm{sync}} \approx 12\times 10^6 (B/\mu \textrm{G})^{-2} (E_e/\rm{TeV})^{-1}\, \rm{yr}
\label{eqn:t_sync}
\end{equation}

where $B$ is given by Equation~\ref{eqn:B_crutcher}.

The Bremsstrahlung cooling time $t_{\rm{brem}}$ is given by:

\begin{equation}
t_{\rm{brem}} \approx 4\times 10^7 (n/\rm{cm}^{3})^{-1}\,  \rm{yr}
\label{eqn:t_brem}
\end{equation}

where $n$ is the number density for each given component. 

The time, $t_{\rm diff}$, it takes for CRs to diffuse across a given distance, $R$, is given by:

\begin{equation}
t_{\rm diff}=R^2/2D(E)
\label{eqn:t_diff}
\end{equation}

where $D(E)$ is the diffusion coefficient (given by Equation~\ref{eqn:diff_coefficient} for $\chi=0.1$) for particles of energy, $E$.

The cooling time for inverse-Compton scattering ($t_{\rm{IC}}$) is estimated to be $230$\,kyr for all ISM components, as it is independent of the ISM density.
The various cooling times for the synchrotron and Bremsstrahlung processes, magnetic field (Equation~\ref{eqn:B_crutcher}), diffusion coefficient (Equation \ref{eqn:diff_coefficient}) and diffusion times (Equation~\ref{eqn:t_diff}) for each gas component are displayed in Table~\ref{tab:cooling}.

\begin{table*}[h]
\begin{center}
\caption{Cooling times for synchrotron radiation, $t_{\rm{sync}}$ (Equation~\ref{eqn:t_sync}), and Bremsstrahlung, $t_{\rm{brem}}$ (Equation~\ref{eqn:t_brem}), towards \hessj for each velocity component defined in Figure~\ref{fig:velSpec}. The diffusion coefficient, $D(E)$, is calculated using Equation~\ref{eqn:diff_coefficient} with use of the magnetic field strength, $B$, within each component. The diffusion time, $t_{\rm{diff}}$, for particles to cross the 30\,pc distance (from \psr to the nearby cloud in component C), is also shown here.}
\begin{tabular}{ccccccc}
\hline
Component & $B$     & $t_{\rm{IC}}$ & $t_{\rm{sync}}$ & $t_{\rm{brem}}$ & $D(E)$     & $t_{\rm{diff}}$ \\
{}       & ($\mu$G) &  (kyr)        & (kyr)           &     (kyr)       & ($10^{27}$\,\diff) & (kyr) \\
\hline
A         &   43    &     230       &    1.3          &     14          & 5.6              & 24 \\
B         &   57    &     230       &    0.73         &     9           & 4.9              & 28 \\
C         &   11    &     230       &    21.6         &     123         & 11               & 12 \\
D         &   10    &     230       &    24.0         &     253         & 12               & 12 \\
E         &   10    &     230       &    24.0         &     1620        & 12               & 12 \\
F         &   10    &     230       &    24.0         &     7120        & 12               & 12 \\
\hline
\end{tabular}
\label{tab:cooling}
\end{center}
\end{table*}

Referring to Table~\ref{tab:cooling}, component C has a magnetic field value of $B=11\mu$G and diffusion coefficient of $D(E)=1.1\times 10^{28}$\,\diff, with a corresponding diffusion time of $12 \,\rm{kyr}$ for electrons to cross the TeV source.

As the pulsar's age ($16$\,kyr) is much less than each of the cooling times, the energy losses from each of the cooling effects are negligible at this stage in the pulsar's life. The diffusion time (Equation~\ref{eqn:t_diff}) for CR electrons of $12$\,kyr is similar to the age of \psr, suggesting electrons are able to diffuse the required distance of 30\,pc in order to contribute to the leptonic TeV emission from \hessj. Therefore the leptonic scenario cannot be ruled out and the spatial extent of the emission is limited by diffusion.

\paragraph{\psrn powered PWN}
The spin down power $6.41\times10^{35}$\,erg\,s$^{-1}$ for \psrn and TeV luminosity of $8.45\times10^{33}$\,erg\,s$^{-1}$ at 1.3\,kpc gives a TeV $\gamma$-ray efficiency of 1\% for \psrn. Therefore, it is possible that \psrn could contribute to the TeV $\gamma$-ray emission from \hessj.

Figure~6 from \cite{PSR_Abdo_2010} shows the population of pulsars with their given $\gamma$-ray luminosity $L_{\gamma}$ and spin down power $\dot{E}$. There is a spread to the data, allowing the authors to place upper ($L_\gamma=\dot{E}$) and lower ($L_\gamma \propto\dot{E}^{1/2}$) bands to this figure. Here the $\gamma$-ray luminosity is given by: 

\begin{equation}
L_\gamma\equiv4\pi d^2 f_{\Omega} \rm{G} _{100}\,  \rm{erg\,s^{-1}}
\label{eqn:lum_g}
\end{equation}

where $f_{\Omega}$ is the flux correction factor set equal to 1 and G$_{100}=13.1\times10^{-11}\,\rm{erg\,cm^{-2}\,s^{-1}}$ is the energy flux obtained from \cite{PSRn_Pletsch_2012}. Equation~\ref{eqn:lum_g} can constrain the distance to \psrn. The lower and upper limits lead to distances of 1.3\,kpc and 6.3\,kpc respectively. As this is within the distances to other counterparts, it is possible that \psrn could be associated with \hessj. The large angular offset between the TeV peak of \hessj and the best-fit position of \psrn of $\mathord{\sim}0.37^{\circ}$ indicates that a PWN scenario seems unlikely. More detailed investigation is however required to understand if \psrn is a viable counterpart to power the source.

\section{CONCLUSION}
In this paper molecular ISM data from the Mopra radio telescope and HI data from the SGPS were used to study the interstellar gas towards the mysterious unidentified TeV $\gamma$-ray source \hessj. CO(1-0) observations showed different velocity components along the line of sight of \hessj which were used to define intriguing features of the interstellar gas along with morphological matches with the TeV $\gamma$-ray emission.

The ISM mass and density derived from the total column density maps were used to test the validity of both the purely hadronic and purely leptonic scenarios for the potential CR accelerators towards \hessj.
Components C, D and C+D were found to contain the bulk of the gas emission towards \hessj. 
Component C shows morphological matches between the $^{12}$CO and TeV gamma-ray emission. There is also a depletion of gas which anti-correlates with the southern TeV peak. Dense gas emission overlaps both \snrg and \hessj in Component D. The addition of components C and D shows an interesting gas feature which follows the outer most contours of \hessj to the south. The southern region of the TeV peak contains a void of gas in this component (C+D). 

For the purely hadronic scenario, \snrg was assumed to be the accelerator of CRs. Sufficient target material for CRs is present in component D (\vlsr=\ 26\ to\ 56\,\kms), corresponding to the distance of \snrg. A total energy budget of $W_{p,\rm{TeV}}\mathord{\sim}1.1\times10^{48}$\,erg for CRs is required, as calculated from the mass of the total target material.
For this scenario we assume CRs have propagated a distance of $R\mathord{\sim}12$\,pc from the accelerator to the cloud, within the lifetime of the SNR. Modelling of the CR spectra showed that the CR interpretation requires slow diffusion ($\chi \le 0.01$) in order to match the observed GeV and TeV CR enhancement factors. It is therefore possible for \snrg to generate the TeV $\gamma$-ray emission from \hessj for the hadronic scenario. 
We also consider CRs being produced from the undetected progenitor SNR of \psr for the hadronic scenario. The derived CR enhancement factors for \hessj and \fges are well matched for $\chi=0.01$ or 0.001.

For the purely leptonic scenario, the TeV emission is produced by highly energetic electrons from \psr as a PWN. A TeV $\gamma$-ray efficiency of $\mathord{\sim}3\%$ was found, supporting this scenario from an energetics point of view. As the diffusion time for CR electrons of 12\,kyr is less than the age of \psr (16\,kyr), the electrons are able to diffuse 30\,pc to create a TeV source of this size. Component C (corresponding to the distance of \psr) shows gas structures which anti-correlate with the TeV emission from \hessj, typical of a PWN driven TeV source. A PWN from \psr could therefore potentially contribute to the TeV $\gamma$-ray emission, so the leptonic scenario cannot be ruled out.

\psrn is also considered for the leptonic scenario. The TeV luminosity at the distance to this pulsar, 1.3\,kpc, requires a 1\% conversion efficiency of the spin-down power of \psrn, a value within the typical efficiencies seen in other firmly identified PWN. However, the large offset between \psrn and the TeV peak of \hessj indicates a PWN scenario is unlikely.

\hessj still remains unidentified in nature due to the complex environment of the initial detection, however, a middle aged SNR or PSR provide a valid interpretation. It may also be possible that the TeV emission has contributions from both leptonic and hadronic processes.
Future work will focus on modelling the spectral energy distribution in more detail, in particular for the case of high-energy electrons.
Future $\gamma$-ray observations from the next-generation ground-based observatory, the Cherenkov Telescope Array (CTA), will provide improved angular resolution (few arcmin) and sensitivity compared to the currently operating telescope arrays. 
These will provide a more detailed look into many unidentified $\gamma$-ray sources, including \hessj, allowing us to further constrain the nature of \hessj.

\begin{acknowledgements}
The Mopra radio telescope is part of the ATNF which is funded by the Australian Government for operation as a National Facility managed by CSIRO (Commonwealth Scientific and Industrial Research Organisation). Support for observations are provided by the University of New South Wales and the University of Adelaide.  This research has made use of the NASA's Astrophysics Data System and the SIMBAD database, operated at CDS, Strasbourg, France. K.F. acknowledges support through the provision of Australian Government Research Training Program Scholarship.

\end{acknowledgements}

\bibliographystyle{pasa-mnras}
\bibliography{pasa-refs}

\begin{thebibliography}{}
\makeatletter
\relax
\def\mn@urlcharsother{\let\do\@makeother \do\$\do\&\do\#\do\^\do\_\do\%\do\~}
\definecolor{darkblue}{rgb}{0,0,0.597656}
\def\mndoi{\begingroup\mn@urlcharsother \@ifnextchar [ {\mndoi@} {\mndoi@[]}}
\def\mndoi@[#1]#2{\def\@tempa{#1}\ifx\@tempa\@empty \href
  {http://dx.doi.org/#2} {\textcolor{darkblue}{doi:#2}}\else \href
  {http://dx.doi.org/#2} {\textcolor{darkblue}{#1}}\fi \endgroup}
\def\mn@eprint#1#2{\mn@eprint@#1:#2::\@nil}
\def\mn@eprint@arXiv#1{\href {http://arxiv.org/abs/#1} {{\tt arXiv:#1}}}
\def\mn@eprint@dblp#1{\href {http://dblp.uni-trier.de/rec/bibtex/#1.xml}
  {dblp:#1}}
\def\mn@eprint@#1:#2:#3:#4\@nil{\def\@tempa {#1}\def\@tempb {#2}\def\@tempc
  {#3}\ifx \@tempc \@empty \let \@tempc \@tempb \let \@tempb \@tempa \fi \ifx
  \@tempb \@empty \def\@tempb {arXiv}\fi \@ifundefined
  {mn@eprint@\@tempb}{\@tempb:\@tempc}{\expandafter \expandafter \csname
  mn@eprint@\@tempb\endcsname \expandafter{\@tempc}}}

\bibitem[\protect\citeauthoryear{{Abdo} et~al.,}{{Abdo}
  et~al.}{2010}]{PSR_Abdo_2010}
{Abdo} A.~A.,  et~al., 2010, \mndoi [\apjs] {10.1088/0067-0049/187/2/460},
  \href {https://ui.adsabs.harvard.edu/abs/2010ApJS..187..460A} {187, 460}

\bibitem[\protect\citeauthoryear{{Abdo} et~al.,}{{Abdo}
  et~al.}{2013}]{PSRn_Abdo_2013}
{Abdo} A.~A.,  et~al., 2013, \mndoi [\apjs] {10.1088/0067-0049/208/2/17}, \href
  {https://ui.adsabs.harvard.edu/abs/2013ApJS..208...17A} {208, 17}

\bibitem[\protect\citeauthoryear{{Acero} et~al.,}{{Acero}
  et~al.}{2016}]{Fermi1_Acero_2016}
{Acero} F.,  et~al., 2016, \mndoi [\apjs] {10.3847/0067-0049/224/1/8}, \href
  {https://ui.adsabs.harvard.edu/abs/2016ApJS..224....8A} {224, 8}

\bibitem[\protect\citeauthoryear{{Ackermann} et~al.,}{{Ackermann}
  et~al.}{2017}]{Fermi_Ackermann_2017}
{Ackermann} M.,  et~al., 2017, \mndoi [\apj] {10.3847/1538-4357/aa775a}, \href
  {https://ui.adsabs.harvard.edu/abs/2017ApJ...843..139A} {843, 139}

\bibitem[\protect\citeauthoryear{Ade et~al.,}{Ade
  et~al.}{2011}]{Eq_Planck_2011}
Ade P.,  et~al., 2011, \mndoi [Astronomy & Astrophysics]
  {10.1051/0004-6361/201116479}, 536

\bibitem[\protect\citeauthoryear{{Aharonian}}{{Aharonian}}{1991}]{VHE_Aharonian_1991}
{Aharonian} F.~A.,  1991, \mndoi [\apss] {10.1007/BF00648185}, \href
  {https://ui.adsabs.harvard.edu/abs/1991Ap&SS.180..305A} {180, 305}

\bibitem[\protect\citeauthoryear{{Aharonian} \& {Atoyan}}{{Aharonian} \&
  {Atoyan}}{1996}]{Pi0decay_Aharonian_1996}
{Aharonian} F.~A.,  {Atoyan} A.~M.,  1996, \aap, \href
  {https://ui.adsabs.harvard.edu/abs/1996A&A...309..917A} {309, 917}

\bibitem[\protect\citeauthoryear{{Aharonian} et~al.,}{{Aharonian}
  et~al.}{2005}]{VHE_Aharonian_2005}
{Aharonian} F.,  et~al., 2005, \mndoi [Science] {10.1126/science.1108643},
  \href {https://ui.adsabs.harvard.edu/abs/2005Sci...307.1938A} {307, 1938}

\bibitem[\protect\citeauthoryear{{Aharonian} et~al.,}{{Aharonian}
  et~al.}{2006}]{HESS_Aharonian_2006}
{Aharonian} F.,  et~al., 2006, \mndoi [\apj] {10.1086/498013}, \href
  {https://ui.adsabs.harvard.edu/abs/2006ApJ...636..777A} {636, 777}

\bibitem[\protect\citeauthoryear{{Ajello} et~al.,}{{Ajello}
  et~al.}{2012}]{Fermi_SNRG_2012}
{Ajello} M.,  et~al., 2012, \mndoi [\apj] {10.1088/0004-637X/744/1/80}, \href
  {https://ui.adsabs.harvard.edu/abs/2012ApJ...744...80A} {744, 80}

\bibitem[\protect\citeauthoryear{{Anderson}, {Bania}, {Balser}, {Cunningham},
  {Wenger}, {Johnstone}  \& {Armentrout}}{{Anderson}
  et~al.}{2014a}]{HIICat_Anderson_2014}
{Anderson} L.~D.,  {Bania} T.~M.,  {Balser} D.~S.,  {Cunningham} V.,  {Wenger}
  T.~V.,  {Johnstone} B.~M.,   {Armentrout} W.~P.,  2014a, VizieR Online Data
  Catalog, \href {https://ui.adsabs.harvard.edu/abs/2014yCat..22120001A} {p.
  J/ApJS/212/1}

\bibitem[\protect\citeauthoryear{{Anderson}, {Bania}, {Balser}
  et~al.}{{Anderson} et~al.}{2014b}]{WISE_Anderson_2014}
{Anderson} L.~D.,  {Bania} T.~M.,  {Balser} D.~S.,   et~al., 2014b, \mndoi
  [APJS] {10.1088/0067-0049/212/1/1}, \href
  {https://ui.adsabs.harvard.edu/abs/2014ApJS..212....1A} {212, 1}

\bibitem[\protect\citeauthoryear{{Berezinskii}, {Bulanov}, {Dogiel}  \&
  {Ptuskin}}{{Berezinskii} et~al.}{1990}]{Book_Berezinskii_1990}
{Berezinskii} V.~S.,  {Bulanov} S.~V.,  {Dogiel} V.~A.,   {Ptuskin} V.~S.,
  1990, {Astrophysics of cosmic rays}

\bibitem[\protect\citeauthoryear{{Bolatto}, {Wolfire}  \& {Leroy}}{{Bolatto}
  et~al.}{2013}]{COdens_Bolatto_2013}
{Bolatto} A.~D.,  {Wolfire} M.,   {Leroy} A.~K.,  2013, \mndoi [\araa]
  {10.1146/annurev-astro-082812-140944}, \href
  {https://ui.adsabs.harvard.edu/abs/2013ARA&A..51..207B} {51, 207}

\bibitem[\protect\citeauthoryear{{Braiding} et~al.,}{{Braiding}
  et~al.}{2018}]{Mopra_Braiding_2018}
{Braiding} C.,  et~al., 2018, \mndoi [\pasa] {10.1017/pasa.2018.18}, \href
  {https://ui.adsabs.harvard.edu/abs/2018PASA...35...29B} {35, e029}

\bibitem[\protect\citeauthoryear{{Brand} \& {Blitz}}{{Brand} \&
  {Blitz}}{1993}]{GRC_Brand_1993}
{Brand} J.,  {Blitz} L.,  1993, \aap, \href
  {https://ui.adsabs.harvard.edu/abs/1993A&A...275...67B} {275, 67}

\bibitem[\protect\citeauthoryear{{Brisken}, {Carrillo-Barrag{\'a}n}, {Kurtz}
  \& {Finley}}{{Brisken} et~al.}{2006}]{PSRJ_Brisken_2006}
{Brisken} W.~F.,  {Carrillo-Barrag{\'a}n} M.,  {Kurtz} S.,   {Finley} J.~P.,
  2006, \mndoi [\apj] {10.1086/507765}, \href
  {https://ui.adsabs.harvard.edu/abs/2006ApJ...652..554B} {652, 554}

\bibitem[\protect\citeauthoryear{{Burton} et~al.,}{{Burton}
  et~al.}{2013}]{Mopra_Burton_2013}
{Burton} M.~G.,  et~al., 2013, \mndoi [\pasa] {10.1017/pasa.2013.22}, \href
  {https://ui.adsabs.harvard.edu/abs/2013PASA...30...44B} {30, e044}

\bibitem[\protect\citeauthoryear{{Casanova} et~al.,}{{Casanova}
  et~al.}{2010}]{tesc_Casanova_2010}
{Casanova} S.,  et~al., 2010, \mndoi [\pasj] {10.1093/pasj/62.5.1127}, \href
  {https://ui.adsabs.harvard.edu/abs/2010PASJ...62.1127C} {62, 1127}

\bibitem[\protect\citeauthoryear{{Claussen}, {Frail}, {Goss}  \&
  {Gaume}}{{Claussen} et~al.}{1997}]{MaserMC_Claussen_1997}
{Claussen} M.~J.,  {Frail} D.~A.,  {Goss} W.~M.,   {Gaume} R.~A.,  1997, \mndoi
  [\apj] {10.1086/304784}, \href
  {https://ui.adsabs.harvard.edu/abs/1997ApJ...489..143C} {489, 143}

\bibitem[\protect\citeauthoryear{{Clifton} \& {Lyne}}{{Clifton} \&
  {Lyne}}{1986}]{PSRs_Clifton_1986}
{Clifton} T.~R.,  {Lyne} A.~G.,  1986, \mndoi [\nat] {10.1038/320043a0}, \href
  {https://ui.adsabs.harvard.edu/abs/1986Natur.320...43C} {320, 43}

\bibitem[\protect\citeauthoryear{{Crutcher}, {Wandelt}, {Heiles}, {Falgarone}
  \& {Troland}}{{Crutcher} et~al.}{2010}]{Mag_Crutcher_2010}
{Crutcher} R.~M.,  {Wandelt} B.,  {Heiles} C.,  {Falgarone} E.,   {Troland}
  T.~H.,  2010, \mndoi [\apj] {10.1088/0004-637X/725/1/466}, \href
  {https://ui.adsabs.harvard.edu/abs/2010ApJ...725..466C} {725, 466}

\bibitem[\protect\citeauthoryear{{Dame}, {Hartmann}  \& {Thaddeus}}{{Dame}
  et~al.}{2001}]{ISM_Dame_2001}
{Dame} T.~M.,  {Hartmann} D.,   {Thaddeus} P.,  2001, \mndoi [\apj]
  {10.1086/318388}, \href
  {https://ui.adsabs.harvard.edu/abs/2001ApJ...547..792D} {547, 792}

\bibitem[\protect\citeauthoryear{{Dermer}}{{Dermer}}{1986}]{EarthCR_Dermer_1986}
{Dermer} C.~D.,  1986, \aap, \href
  {https://ui.adsabs.harvard.edu/abs/1986A%26A...157..223D} {157, 223}

\bibitem[\protect\citeauthoryear{{Dickey} \& {Lockman}}{{Dickey} \&
  {Lockman}}{1990}]{HI_Dickey_1990}
{Dickey} J.~M.,  {Lockman} F.~J.,  1990, \mndoi [\araa]
  {10.1146/annurev.aa.28.090190.001243}, \href
  {https://ui.adsabs.harvard.edu/abs/1990ARA&A..28..215D} {28, 215}

\bibitem[\protect\citeauthoryear{{Fang} \& {Zhang}}{{Fang} \&
  {Zhang}}{2008}]{Old_SNR_Fang_2008}
{Fang} J.,  {Zhang} L.,  2008, \mndoi [\mnras]
  {10.1111/j.1365-2966.2007.12766.x}, \href
  {https://ui.adsabs.harvard.edu/abs/2008MNRAS.384.1119F} {384, 1119}

\bibitem[\protect\citeauthoryear{{Fernandez}, {Dalton}, {Eger}, {Laffon},
  {Mehault}, {Ohm}, {Oya}  \& {M.~Renaud for the H.E.S.S.
  Collaboration}}{{Fernandez} et~al.}{2013}]{SNR_MCs_Fernandez_2013}
{Fernandez} D.,  {Dalton} M.,  {Eger} P.,  {Laffon} H.,  {Mehault} J.,  {Ohm}
  S.,  {Oya} I.,   {M.~Renaud for the H.E.S.S. Collaboration} 2013, preprint,
  \href {https://ui.adsabs.harvard.edu/abs/2013arXiv1305.6396F} {} (\mn@eprint
  {arXiv} {1305.6396})

\bibitem[\protect\citeauthoryear{{Finkbeiner}}{{Finkbeiner}}{2003}]{Table_Finkbeiner_2003}
{Finkbeiner} D.~P.,  2003, \mndoi [\apjs] {10.1086/374411}, \href
  {https://ui.adsabs.harvard.edu/abs/2003ApJS..146..407F} {146, 407}

\bibitem[\protect\citeauthoryear{{Finley} \& {Oegelman}}{{Finley} \&
  {Oegelman}}{1994}]{SNRage_Finley_1994}
{Finley} J.~P.,  {Oegelman} H.,  1994, \mndoi [\apjl] {10.1086/187563}, \href
  {https://ui.adsabs.harvard.edu/abs/1994ApJ...434L..25F} {434, L25}

\bibitem[\protect\citeauthoryear{{Frail}, {Goss}  \& {Slysh}}{{Frail}
  et~al.}{1994}]{W28_Frail_1994}
{Frail} D.~A.,  {Goss} W.~M.,   {Slysh} V.~I.,  1994, \mndoi [\apjl]
  {10.1086/187287}, \href
  {https://ui.adsabs.harvard.edu/abs/1994ApJ...424L.111F} {424, L111}

\bibitem[\protect\citeauthoryear{{Frerking}, {Wilson}, {Linke}  \&
  {Wannier}}{{Frerking} et~al.}{1980}]{CS_Frerking_1980}
{Frerking} M.~A.,  {Wilson} R.~W.,  {Linke} R.~A.,   {Wannier} P.~G.,  1980,
  \mndoi [\apj] {10.1086/158207}, \href
  {https://ui.adsabs.harvard.edu/abs/1980ApJ...240...65F} {240, 65}

\bibitem[\protect\citeauthoryear{{Gabici}, {Aharonian}  \& {Blasi}}{{Gabici}
  et~al.}{2007}]{Lep_Gabici_2007}
{Gabici} S.,  {Aharonian} F.~A.,   {Blasi} P.,  2007, \mndoi [\apss]
  {10.1007/s10509-007-9427-6}, \href
  {https://ui.adsabs.harvard.edu/abs/2007Ap&SS.309..365G} {309, 365}

\bibitem[\protect\citeauthoryear{{Gabici}, {Aharonian}  \& {Casanova}}{{Gabici}
  et~al.}{2009}]{Gabici_2009}
{Gabici} S.,  {Aharonian} F.~A.,   {Casanova} S.,  2009, \mndoi [\mnras]
  {10.1111/j.1365-2966.2009.14832.x}, \href
  {https://ui.adsabs.harvard.edu/abs/2009MNRAS.396.1629G} {396, 1629}

\bibitem[\protect\citeauthoryear{{Gabici}, {Casanova}, {Aharonian}  \&
  {Rowell}}{{Gabici} et~al.}{2010}]{chi_Gabici_2010}
{Gabici} S.,  {Casanova} S.,  {Aharonian} F.~A.,   {Rowell} G.,  2010, in
  {Boissier} S.,  {Heydari-Malayeri} M.,  {Samadi} R.,   {Valls-Gabaud} D.,
  eds, SF2A-2010: Proceedings of the Annual meeting of the French Society of
  Astronomy and Astrophysics. p.~313 (\mn@eprint {arXiv} {1009.5291})

\bibitem[\protect\citeauthoryear{{Giuliani} et~al.,}{{Giuliani}
  et~al.}{2010}]{Chi_Giuliani_2010}
{Giuliani} A.,  et~al., 2010, \mndoi [\aap] {10.1051/0004-6361/201014256},
  \href {https://ui.adsabs.harvard.edu/abs/2010A&A...516L..11G} {516, L11}

\bibitem[\protect\citeauthoryear{{Gusdorf}, {Cabrit}, {Flower}  \& {Pineau Des
  For{\^e}ts}}{{Gusdorf} et~al.}{2008}]{SiO_Gusdorf_2008}
{Gusdorf} A.,  {Cabrit} S.,  {Flower} D.~R.,   {Pineau Des For{\^e}ts} G.,
  2008, \mndoi [\aap] {10.1051/0004-6361:20078900}, \href
  {https://ui.adsabs.harvard.edu/abs/2008A&A...482..809G} {482, 809}

\bibitem[\protect\citeauthoryear{{H.E.S.S. Collaboration} et~al.,}{{H.E.S.S.
  Collaboration} et~al.}{2018a}]{HGPS_2018}
{H.E.S.S. Collaboration} et~al., 2018a, \mndoi [\aap]
  {10.1051/0004-6361/201732098}, \href
  {https://ui.adsabs.harvard.edu/abs/2018A&A...612A...1H} {612, A1}

\bibitem[\protect\citeauthoryear{{H.E.S.S. Collaboration} et~al.,}{{H.E.S.S.
  Collaboration} et~al.}{2018b}]{PWNe_Aharonian_2018}
{H.E.S.S. Collaboration} et~al., 2018b, \mndoi [\aap]
  {10.1051/0004-6361/201629377}, \href
  {https://ui.adsabs.harvard.edu/abs/2018A&A...612A...2H} {612, A2}

\bibitem[\protect\citeauthoryear{{Hewitt} \& {Yusef-Zadeh}}{{Hewitt} \&
  {Yusef-Zadeh}}{2009}]{SNRs_Hewitt_2009}
{Hewitt} J.~W.,  {Yusef-Zadeh} F.,  2009, \mndoi [\apjl]
  {10.1088/0004-637X/694/1/L16}, \href
  {https://ui.adsabs.harvard.edu/abs/2009ApJ...694L..16H} {694, L16}

\bibitem[\protect\citeauthoryear{{Higashi} et~al.,}{{Higashi}
  et~al.}{2008}]{HESSJ_Higashi_2008}
{Higashi} Y.,  et~al., 2008, \mndoi [\apj] {10.1086/589877}, \href
  {https://ui.adsabs.harvard.edu/abs/2008ApJ...683..957H} {683, 957}

\bibitem[\protect\citeauthoryear{{Ho} \& {Townes}}{{Ho} \&
  {Townes}}{1983}]{NH3_Ho_1983}
{Ho} P.~T.~P.,  {Townes} C.~H.,  1983, \mndoi [\araa]
  {10.1146/annurev.aa.21.090183.001323}, \href
  {https://ui.adsabs.harvard.edu/abs/1983ARA&A..21..239H} {21, 239}

\bibitem[\protect\citeauthoryear{{Jackson} et~al.,}{{Jackson}
  et~al.}{2013}]{HC3N_2013}
{Jackson} J.~M.,  et~al., 2013, \mndoi [\pasa] {10.1017/pasa.2013.37}, \href
  {https://ui.adsabs.harvard.edu/abs/2013PASA...30...57J} {30, e057}

\bibitem[\protect\citeauthoryear{{Kargaltsev}, {Pavlov}  \&
  {Garmire}}{{Kargaltsev} et~al.}{2007a}]{Xrays_Kargaltsev_2007}
{Kargaltsev} O.,  {Pavlov} G.~G.,   {Garmire} G.~P.,  2007a, \mndoi [\apj]
  {10.1086/513312}, \href
  {https://ui.adsabs.harvard.edu/abs/2007ApJ...660.1413K} {660, 1413}

\bibitem[\protect\citeauthoryear{{Kargaltsev}, {Pavlov}  \&
  {Garmire}}{{Kargaltsev} et~al.}{2007b}]{HESSJ_Karg_2007}
{Kargaltsev} O.,  {Pavlov} G.~G.,   {Garmire} G.~P.,  2007b, \mndoi [APJ]
  {10.1086/521520}, \href
  {https://ui.adsabs.harvard.edu/abs/2007ApJ...670..643K} {670, 643}

\bibitem[\protect\citeauthoryear{{Kassim} \& {Weiler}}{{Kassim} \&
  {Weiler}}{1990}]{W30_Kassim_1990}
{Kassim} N.~E.,  {Weiler} K.~W.,  1990, \mndoi [\apj] {10.1086/169107}, \href
  {https://ui.adsabs.harvard.edu/abs/1990ApJ...360..184K} {360, 184}

\bibitem[\protect\citeauthoryear{{Kelner}, {Aharonian}  \& {Bugayov}}{{Kelner}
  et~al.}{2006}]{X_Kelner_2006}
{Kelner} S.~R.,  {Aharonian} F.~A.,   {Bugayov} V.~V.,  2006, \mndoi [\prd]
  {10.1103/PhysRevD.74.034018}, \href
  {https://ui.adsabs.harvard.edu/abs/2006PhRvD..74c4018K} {74, 034018}

\bibitem[\protect\citeauthoryear{{Kilpatrick}, {Bieging}  \&
  {Rieke}}{{Kilpatrick} et~al.}{2016}]{CO_Kilpatrick_2016}
{Kilpatrick} C.~D.,  {Bieging} J.~H.,   {Rieke} G.~H.,  2016, \mndoi [\apj]
  {10.3847/0004-637X/816/1/1}, \href
  {https://ui.adsabs.harvard.edu/abs/2016ApJ...816....1K} {816, 1}

\bibitem[\protect\citeauthoryear{{Ladd}, {Purcell}, {Wong}  \&
  {Robertson}}{{Ladd} et~al.}{2005}]{ATNF_Ladd_2005}
{Ladd} N.,  {Purcell} C.,  {Wong} T.,   {Robertson} S.,  2005, \mndoi [\pasa]
  {10.1071/AS04068}, \href
  {https://ui.adsabs.harvard.edu/abs/2005PASA...22...62L} {22, 62}

\bibitem[\protect\citeauthoryear{Li \& Chen}{Li \& Chen}{2010}]{Chi_Hui_2010}
Li H.,  Chen Y.,  2010, \mndoi [Monthly Notices of the Royal Astronomical
  Society: Letters] {10.1111/j.1745-3933.2010.00944.x}, 409, L35

\bibitem[\protect\citeauthoryear{Li \& Chen}{Li \& Chen}{2012}]{Chi_Hui_2012}
Li H.,  Chen Y.,  2012, \mndoi [Monthly Notices of the Royal Astronomical
  Society] {10.1111/j.1365-2966.2012.20270.x}, 421, 935

\bibitem[\protect\citeauthoryear{{Li} et~al.,}{{Li}
  et~al.}{2018}]{Dark_Li_2018}
{Li} D.,  et~al., 2018, \mndoi [\apjs] {10.3847/1538-4365/aaa762}, \href
  {https://ui.adsabs.harvard.edu/abs/2018ApJS..235....1L} {235, 1}

\bibitem[\protect\citeauthoryear{{Lin}, {Webb}  \& {Barret}}{{Lin}
  et~al.}{2013}]{HESSJ_Lin_2013}
{Lin} D.,  {Webb} N.~A.,   {Barret} D.,  2013, \mndoi [APJ]
  {10.1088/0004-637X/766/1/29}, \href
  {https://ui.adsabs.harvard.edu/abs/2013ApJ...766...29L} {766, 29}

\bibitem[\protect\citeauthoryear{{Martin-Pintado}, {Bachiller}  \&
  {Fuente}}{{Martin-Pintado} et~al.}{1992}]{SiO_Martin_1992}
{Martin-Pintado} J.,  {Bachiller} R.,   {Fuente} A.,  1992, \aap, \href
  {https://ui.adsabs.harvard.edu/abs/1992A&A...254..315M} {254, 315}

\bibitem[\protect\citeauthoryear{{McClure-Griffiths}, {Dickey}, {Gaensler},
  {Green}, {Haverkorn}  \& {Strasser}}{{McClure-Griffiths}
  et~al.}{2005}]{SGPS_Naomi_2005}
{McClure-Griffiths} N.~M.,  {Dickey} J.~M.,  {Gaensler} B.~M.,  {Green} A.~J.,
  {Haverkorn} M.,   {Strasser} S.,  2005, \mndoi [\apjs] {10.1086/430114},
  \href {https://ui.adsabs.harvard.edu/abs/2005ApJS..158..178M} {158, 178}

\bibitem[\protect\citeauthoryear{{Morris} et~al.,}{{Morris}
  et~al.}{2002}]{PSRJ1806_Parkes_2002}
{Morris} D.~J.,  et~al., 2002, \mndoi [\mnras]
  {10.1046/j.1365-8711.2002.05551.x}, \href
  {https://ui.adsabs.harvard.edu/abs/2002MNRAS.335..275M} {335, 275}

\bibitem[\protect\citeauthoryear{{Nicholas}, {Rowell}, {Burton}, {Walsh},
  {Fukui}, {Kawamura}  \& {Maxted}}{{Nicholas}
  et~al.}{2012}]{W28_Nicholas_2012}
{Nicholas} B.~P.,  {Rowell} G.,  {Burton} M.~G.,  {Walsh} A.~J.,  {Fukui} Y.,
  {Kawamura} A.,   {Maxted} N.~I.,  2012, \mndoi [\mnras]
  {10.1111/j.1365-2966.2011.19688.x}, \href
  {https://ui.adsabs.harvard.edu/abs/2012MNRAS.419..251N} {419, 251}

\bibitem[\protect\citeauthoryear{{Odegard}}{{Odegard}}{1986}]{SNRage_Odegard_1986}
{Odegard} N.,  1986, \mndoi [\aj] {10.1086/114270}, \href
  {https://ui.adsabs.harvard.edu/abs/1986AJ.....92.1372O} {92, 1372}

\bibitem[\protect\citeauthoryear{{Penzias}, {Solomon}, {Wilson}  \&
  {Jefferts}}{{Penzias} et~al.}{1971}]{CS_Penzias_1971}
{Penzias} A.~A.,  {Solomon} P.~M.,  {Wilson} R.~W.,   {Jefferts} K.~B.,  1971,
  \mndoi [\apjl] {10.1086/180784}, \href
  {https://ui.adsabs.harvard.edu/abs/1971ApJ...168L..53P} {168, L53}

\bibitem[\protect\citeauthoryear{{Planck Collaboration} et~al.,}{{Planck
  Collaboration} et~al.}{2016}]{Planck_collab_2016}
{Planck Collaboration} et~al., 2016, \mndoi [\aap]
  {10.1051/0004-6361/201525967}, \href
  {https://ui.adsabs.harvard.edu/abs/2016A&A...594A..10P} {594, A10}

\bibitem[\protect\citeauthoryear{{Pletsch} et~al.,}{{Pletsch}
  et~al.}{2012}]{PSRn_Pletsch_2012}
{Pletsch} H.~J.,  et~al., 2012, \mndoi [\apj] {10.1088/0004-637X/744/2/105},
  \href {https://ui.adsabs.harvard.edu/abs/2012ApJ...744..105P} {744, 105}

\bibitem[\protect\citeauthoryear{{Protheroe}, {Ott}, {Ekers}, {Jones}  \&
  {Crocker}}{{Protheroe} et~al.}{2008}]{tdiff_Protheroe_2008}
{Protheroe} R.~J.,  {Ott} J.,  {Ekers} R.~D.,  {Jones} D.~I.,   {Crocker}
  R.~M.,  2008, \mndoi [\mnras] {10.1111/j.1365-2966.2008.13752.x}, \href
  {https://ui.adsabs.harvard.edu/abs/2008MNRAS.390..683P} {390, 683}

\bibitem[\protect\citeauthoryear{{Qasim}, {Chuang}, {Fedoseev}, {Ioppolo},
  {Boogert}  \& {Linnartz}}{{Qasim} et~al.}{2018}]{CH3OH_Qasim_2018}
{Qasim} D.,  {Chuang} K.-J.,  {Fedoseev} G.,  {Ioppolo} S.,  {Boogert}
  A.~C.~A.,   {Linnartz} H.,  2018, \mndoi [\aap]
  {10.1051/0004-6361/201732355}, \href
  {https://ui.adsabs.harvard.edu/abs/2018A&A...612A..83Q} {612, A83}

\bibitem[\protect\citeauthoryear{Reynolds}{Reynolds}{2008}]{SNR_Reynolds_2008}
Reynolds S.~P.,  2008, \mndoi [Annual Review of Astronomy and Astrophysics]
  {10.1146/annurev.astro.46.060407.145237}, 46, 89

\bibitem[\protect\citeauthoryear{{Saz Parkinson} et~al.,}{{Saz Parkinson}
  et~al.}{2010}]{PSRn_Dist_2010}
{Saz Parkinson} P.~M.,  et~al., 2010, \mndoi [\apj]
  {10.1088/0004-637X/725/1/571}, \href
  {https://ui.adsabs.harvard.edu/abs/2010ApJ...725..571S} {725, 571}

\bibitem[\protect\citeauthoryear{{Simon}, {Jackson}, {Clemens}, {Bania}  \&
  {Heyer}}{{Simon} et~al.}{2001}]{MC_Simon_2001}
{Simon} R.,  {Jackson} J.~M.,  {Clemens} D.~P.,  {Bania} T.~M.,   {Heyer}
  M.~H.,  2001, \mndoi [\apj] {10.1086/320230}, \href
  {https://ui.adsabs.harvard.edu/abs/2001ApJ...551..747S} {551, 747}

\bibitem[\protect\citeauthoryear{{Sodroski}, {Odegard}, {Arendt}, {Dwek},
  {Weiland}, {Hauser}  \& {Kelsall}}{{Sodroski}
  et~al.}{1997}]{HII_Sodroski_1997}
{Sodroski} T.~J.,  {Odegard} N.,  {Arendt} R.~G.,  {Dwek} E.,  {Weiland} J.~L.,
   {Hauser} M.~G.,   {Kelsall} T.,  1997, \mndoi [\apj] {10.1086/303961}, \href
  {https://ui.adsabs.harvard.edu/abs/1997ApJ...480..173S} {480, 173}

\bibitem[\protect\citeauthoryear{{Urquhart} et~al.,}{{Urquhart}
  et~al.}{2010}]{Mopra_Urquhart_2010}
{Urquhart} J.~S.,  et~al., 2010, \mndoi [\pasa] {10.1071/AS10002}, \href
  {https://ui.adsabs.harvard.edu/abs/2010PASA...27..321U} {27, 321}

\bibitem[\protect\citeauthoryear{{Vall{\'e}e}}{{Vall{\'e}e}}{2014}]{GRC_Vallee_2014}
{Vall{\'e}e} J.~P.,  2014, \mndoi [\aj] {10.1088/0004-6256/148/1/5}, \href
  {https://ui.adsabs.harvard.edu/abs/2014AJ....148....5V} {148, 5}

\bibitem[\protect\citeauthoryear{{Voronkov}, {Caswell}, {Ellingsen}  \&
  {Sobolev}}{{Voronkov} et~al.}{2010}]{CH3OH_Voronkov_2010}
{Voronkov} M.~A.,  {Caswell} J.~L.,  {Ellingsen} S.~P.,   {Sobolev} A.~M.,
  2010, \mndoi [\mnras] {10.1111/j.1365-2966.2010.16624.x}, \href
  {https://ui.adsabs.harvard.edu/abs/2010MNRAS.405.2471V} {405, 2471}

\bibitem[\protect\citeauthoryear{Walsh et~al.,}{Walsh
  et~al.}{2011}]{NH3_Walsh_2011}
Walsh A.~J.,  et~al., 2011, \mndoi [Monthly Notices of the Royal Astronomical
  Society] {10.1111/j.1365-2966.2011.19115.x}, 416, 1764

\bibitem[\protect\citeauthoryear{{Yamazaki}, {Kohri}, {Bamba}, {Yoshida},
  {Tsuribe}  \& {Takahara}}{{Yamazaki} et~al.}{2006}]{Old_SNR_Yamazaki_2006}
{Yamazaki} R.,  {Kohri} K.,  {Bamba} A.,  {Yoshida} T.,  {Tsuribe} T.,
  {Takahara} F.,  2006, \mndoi [\mnras] {10.1111/j.1365-2966.2006.10832.x},
  \href {https://ui.adsabs.harvard.edu/abs/2006MNRAS.371.1975Y} {371, 1975}

\bibitem[\protect\citeauthoryear{{de Wilt}, {Rowell}, {Walsh}, {Burton},
  {Rathborne}, {Fukui}, {Kawamura}  \& {Aharonian}}{{de Wilt}
  et~al.}{2017}]{1804_deWilt_2017}
{de Wilt} P.,  {Rowell} G.,  {Walsh} A.~J.,  {Burton} M.,  {Rathborne} J.,
  {Fukui} Y.,  {Kawamura} A.,   {Aharonian} F.,  2017, \mndoi [\mnras]
  {10.1093/mnras/stx369}, \href
  {https://ui.adsabs.harvard.edu/abs/2017MNRAS.468.2093D} {468, 2093}

\makeatother
\end{thebibliography}

\clearpage

\appendix
\counterwithin{figure}{section}
\counterwithin{equation}{section}
\counterwithin{table}{section}

\section*{APPENDIX}

\section{Pulsar Proper Motion}
\label{subsec:PSR}
The proper motion of \psr has been studied by \cite{PSRJ_Brisken_2006} via radio observations. The proper motion of \psr has been calculated for the Right Ascension (RA) and Declination (Dec), $\mu_{\alpha}=(11.6\pm1.8) \times 10^{-3}$\,arcsec\,yr$^{-1}$ and $\mu_{\delta}=(14.8\pm2.3) \times 10^{-3}$\,arcsec\,yr$^{-1}$ respectively. Given an age of $\mathord{\sim}16$\,kyr for \psr, a birth position for the pulsar of RA=$18^{\rm{h}}03^{\rm{m}}38^{\rm{s}}.0$, Dec=$-21^{\circ}41'18''.2$ is obtained, placing it on edge of the W30 SNR, \snrg.

\begin{figure}[!h]
\includegraphics[width=\columnwidth]{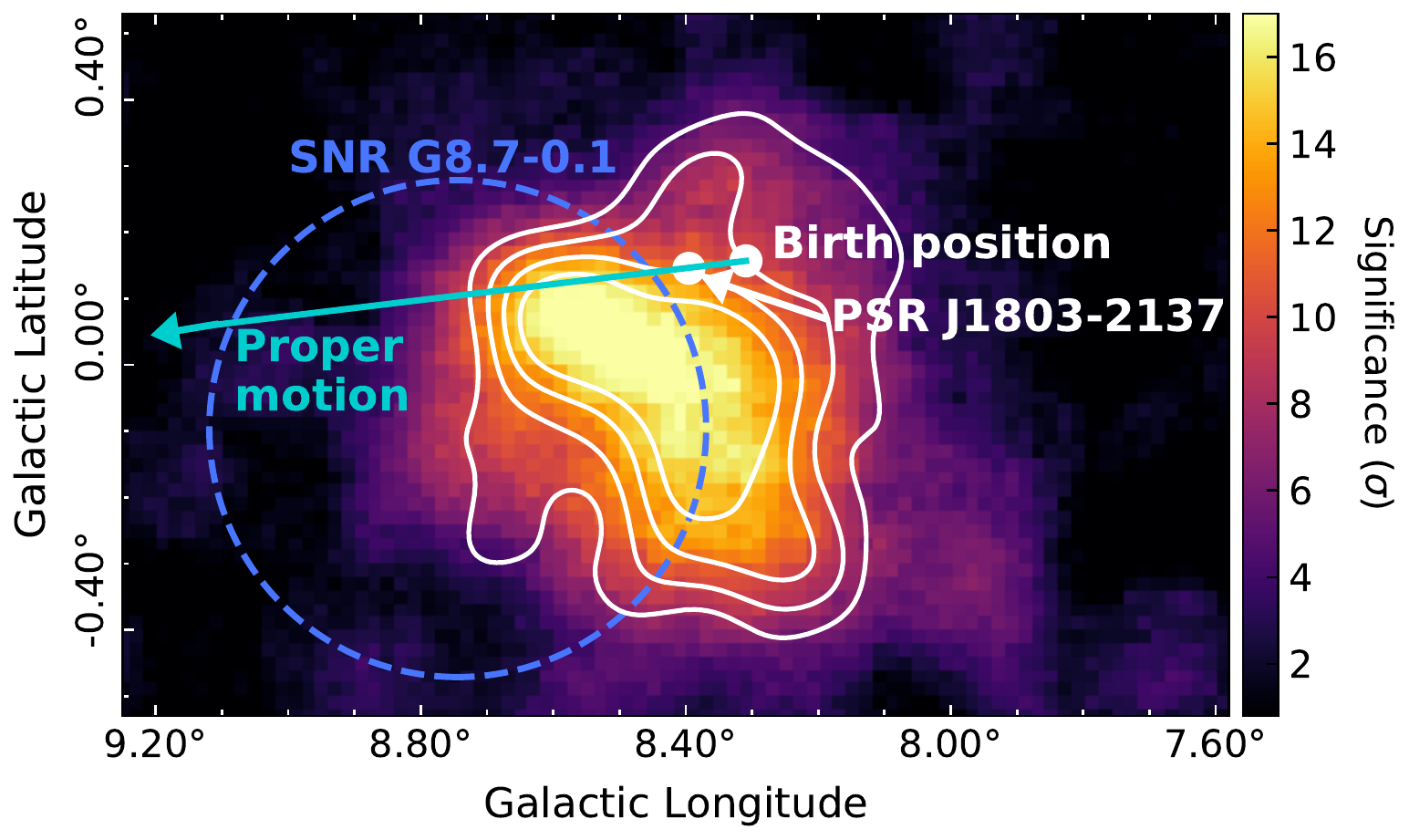}
\caption{TeV $\gamma$-ray significance image of \hessj \citep{HGPS_2018}, showing the proposed proper motion of \psr \citep{PSRJ_Brisken_2006}. The TeV $\gamma$-ray emission for 5-10$\sigma$ is shown by the solid white contours, \snrg is shown by the blue dashed circle, the white dots indicate \psr and its birth position.}
\label{fig:PSR_motion}
\end{figure}

\section{HII regions}
\label{subsec:HII}
\begin{figure}[!h]
\includegraphics[width=\columnwidth]{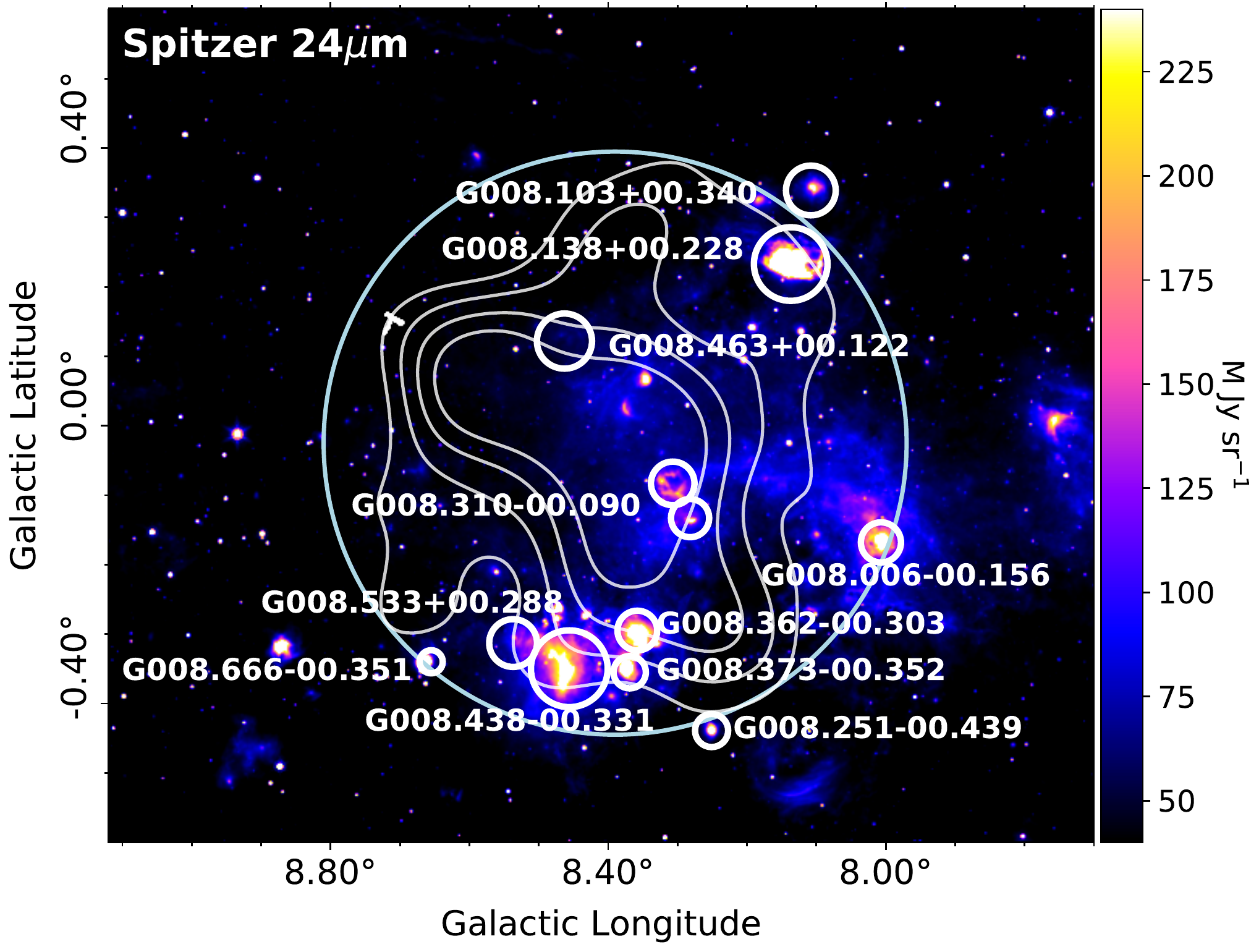}
\caption{$24\,\mu$m infrared image [M\,Jy\,sr$^{-1}$] towards \hessj from the Spitzer GLIMPSE Survey. The TeV $\gamma$-ray emission for 5-10$\sigma$ is shown by the solid white contours, with the cyan circle showing the extent of \hessj. H\textsc{ii} regions with a radius greater than 50$'$ are indicated by the white circles from WISE \citep{WISE_Anderson_2014}.}
\label{fig:HII_reg}
\end{figure}

H\textsc{ii} data was used from the WISE \citep{WISE_Anderson_2014} satellite in order to reveal the known H\textsc{ii} regions towards \hessj (Figure~\ref{fig:HII_reg}). These regions where chosen such that their radius was larger than 50\,arcmins. The online catalogue \citep{HIICat_Anderson_2014} provides the velocity (\vlsr) of each H\textsc{ii} region, which correspond to kinematic distances ranging from 3\,kpc to 5\,kpc. 

\section{Position-velocity plot}
\label{subsec:PV}
\begin{figure}[h!]
\includegraphics[width=\columnwidth]{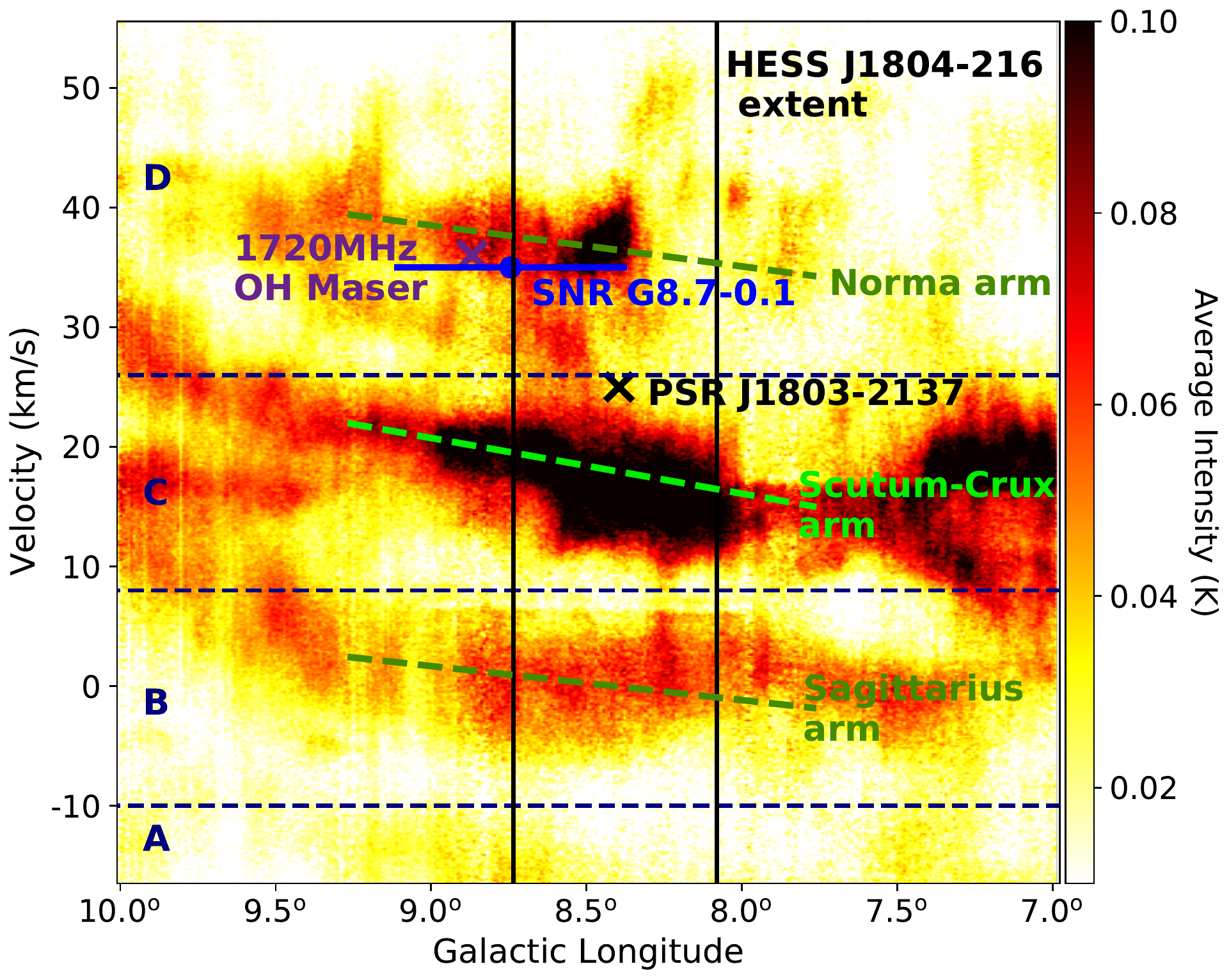}
\caption{Position-velocity plot of Mopra $^{12}$CO(1-0) emission (K) towards \hessj. The black vertical lines show the longitudinal extent of \hessj. The black cross indicates the location of \psr at its assumed velocity of $\mathord{\sim}25$\,\kms. The 1720MHz OH maser is shown by the purple cross at its velocity of 36\,\kms. The centre of \snrg is shown by the blue dot, whilst the blue line shows its radial extent. The green dashed lines are estimates of the Galactic spiral arms along the line of sight for \hessj \citep[from the model in][]{GRC_Vallee_2014}.}
\label{fig:PVplot}
\end{figure}

Figure~\ref{fig:PVplot} is a position-velocity plot towards the \hessj region. This figure shows distinct Mopra $^{12}$CO(1-0) emission in the velocity ranges from \\ \vlsr$=10$\ to\ $25$\,\kms and \vlsr$=35$\ to\ $40$\,\kms which is consistent with the molecular gas discussed in Section~\ref{subsec:ISM_comps}.

\section{Galactic Rotation Curve}
\label{subsec:GRC}
\begin{figure}[h!]
\includegraphics[width=\columnwidth]{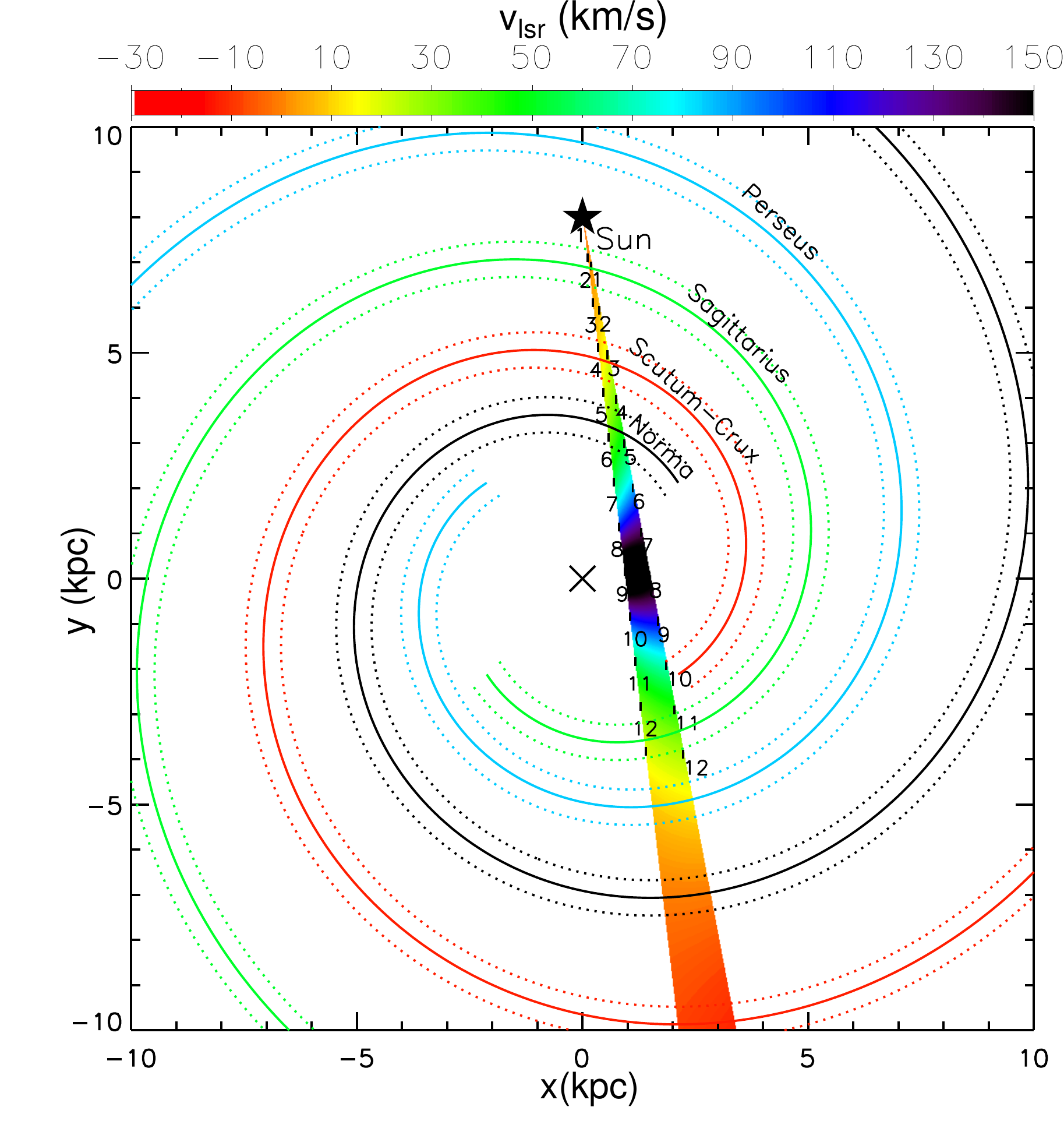}
\caption{Model of the Galaxy along the line of sight of \hessj. Parameters used in this model are from \cite{GRC_Vallee_2014} for each spiral arm shown by the solid coloured lines, Perseus (light blue), Sagittarius (light green), Scutum-Crux (red) and Norma (black). The dashed lines for each spiral arm show their extent. The coloured wedge shows the expected line of sight for \hessj from the Sun for the radial velocities (\vlsr) using the galactic rotation model from \cite{GRC_Brand_1993}. The numbers along this wedge show the distance to the source in kilo-parsecs (kpc). The spatial coordinates along the axes are given in kpc also.}
\label{fig:1804GalRot}
\end{figure}

Objects within the galaxy are rotating around the Galactic Centre (GC). The Galactic rotation curve is a model which gives the average velocity of an object in the galaxy with respect to the GC as a function of distance. The kinematic distance to an object can be found by knowing the position and radial velocity of the given object, from Equation~\ref{eqn:vlsr} \citep{GRC_Brand_1993}. 

\begin{equation}
\mathrm{v}_{\mathrm{lsr}} = \left[ \dfrac{\Theta R_0}{R} - \Theta_0 \right] \sin(l)\cos(b)
\label{eqn:vlsr}
\end{equation}

where $R$ is the galactocentric distance (distance of an object from the centre of the Milky Way galaxy) to the object, $\Theta$ is the circular rotation velocity of object. $R_0$ is the galactocentric distance from the sun and $\Theta_0$ is the circular rotation velocity at the position of the Sun, commonly given values 8.5\,kpc and 220\,\kms respectively. The galactic coordinates are given by $l$ (galactic longitude) and $b$ (galactic latitude). The galactic model along the line of sight for \hessj is shown by Figure~\ref{fig:1804GalRot}. 

\section{HI Spectra and Absorption}
\label{subsec:HI_abs}
\begin{figure}
\begin{center}
\includegraphics[width=\columnwidth]{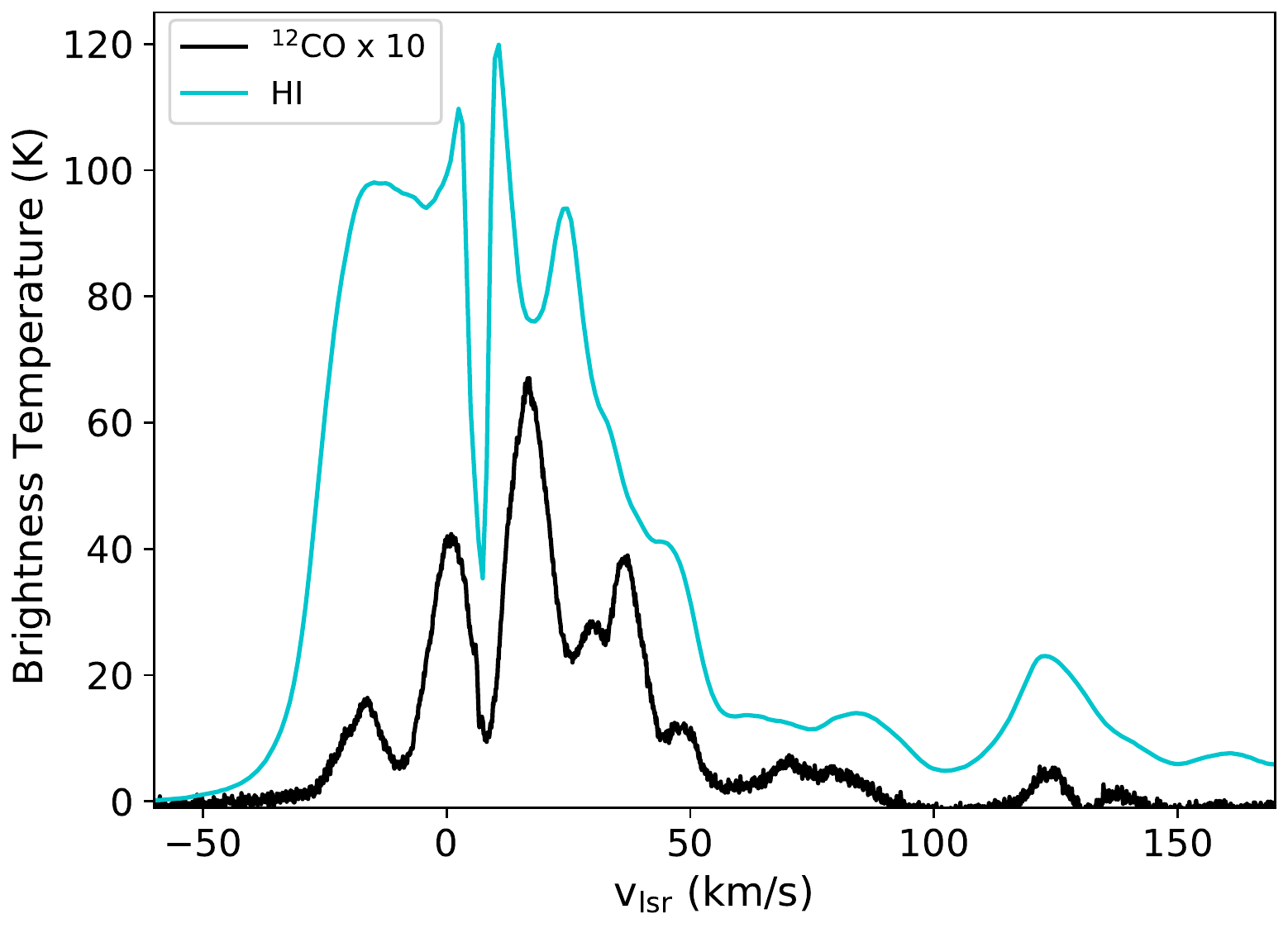}
\caption{Emission spectrum towards \hessj. Solid black lines and cyan lines represent the spectrum for Mopra $^{12}$CO(1-0) and SGPS HI respectively. $^{12}$CO is scaled by a factor of 10 for clarity.}
\label{fig:velSpecHI}
\end{center}
\end{figure}

The spectra of the HI data is shown in Figure~\ref{fig:velSpecHI}. Dips tend to occur in the HI spectra which result from either the presence of a background source that leads to absorption or from HI self-absorption. A well defined HI absorption feature is present at \vlsr$\mathord{\sim}20$\,\kms which corresponds to a strong emission feature in the $^{12}$CO spectra (shown in Figure~\ref{fig:velSpec}). This strong absorption feature could be due to a continuum source, such as \snrg. These properties indicate that the gas is most likely to be foreground to \snrg. This helps to constrain the distance to the SNR, proving that the pre-defined distance of 4.5\,kpc (seen in Section~\ref{sec:intro}) is consistent with the gas data analysis shown here.

\section{Integrated intensity maps}
\label{subsec:maps}

\begin{figure*}
\begin{center}
\includegraphics[width=1.4\columnwidth]{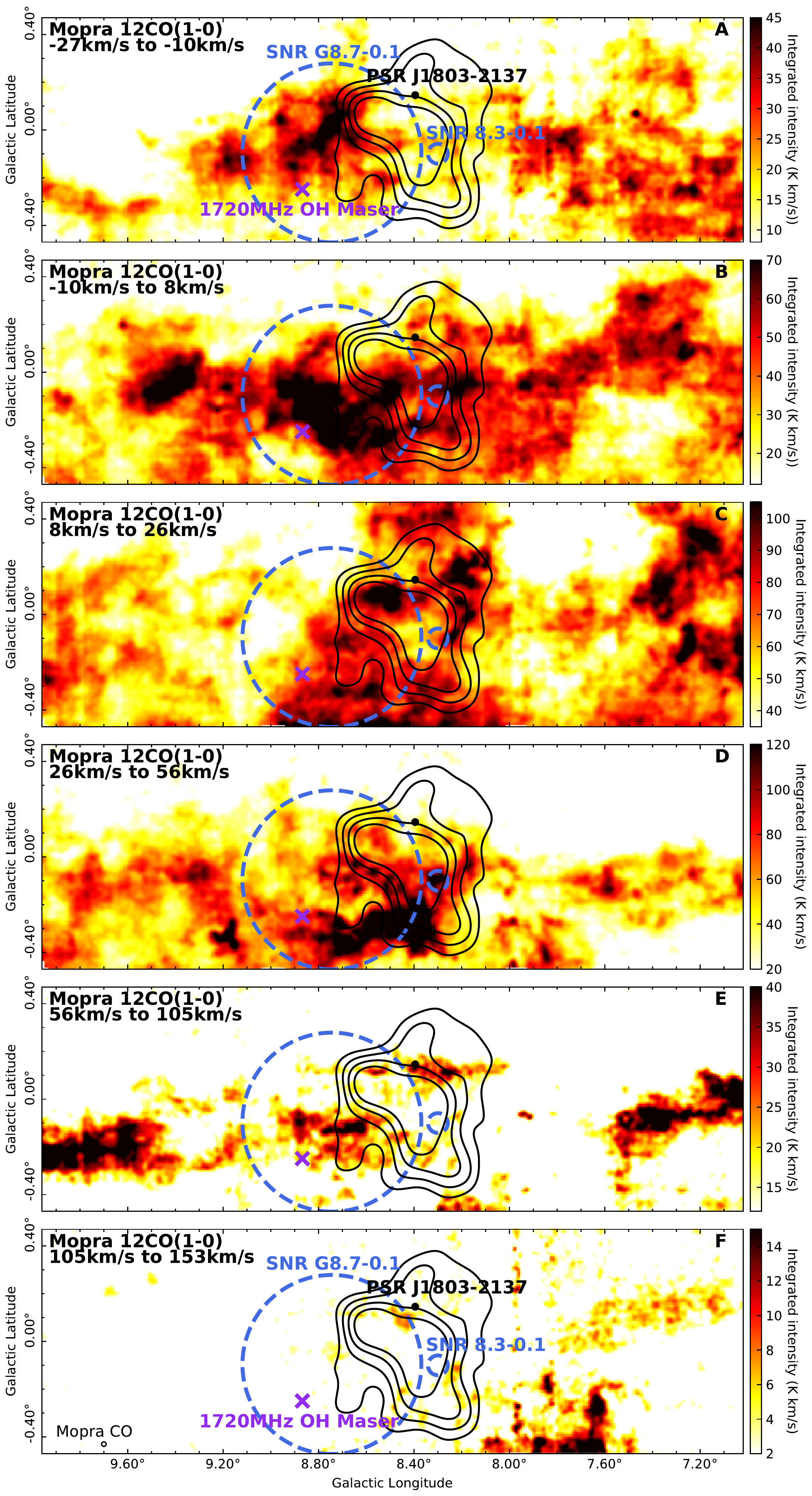}
\caption{Mosaic of Mopra $^{12}$CO integrated intensity maps (K\,\kms) towards \hessj, for gas components A-F as defined in Figure~\ref{fig:velSpec}. The two dashed blue circles indicate \snrg and \snr. The \OH is indicated by the purple cross and \psr is indicated by the black dot. The TeV $\gamma$-ray emission for 5-10$\sigma$ is shown by the solid black contours.}
\label{fig:12CO_mop}
\end{center}
\end{figure*}

\begin{figure*}
\begin{center}
\includegraphics[width=1.4\columnwidth]{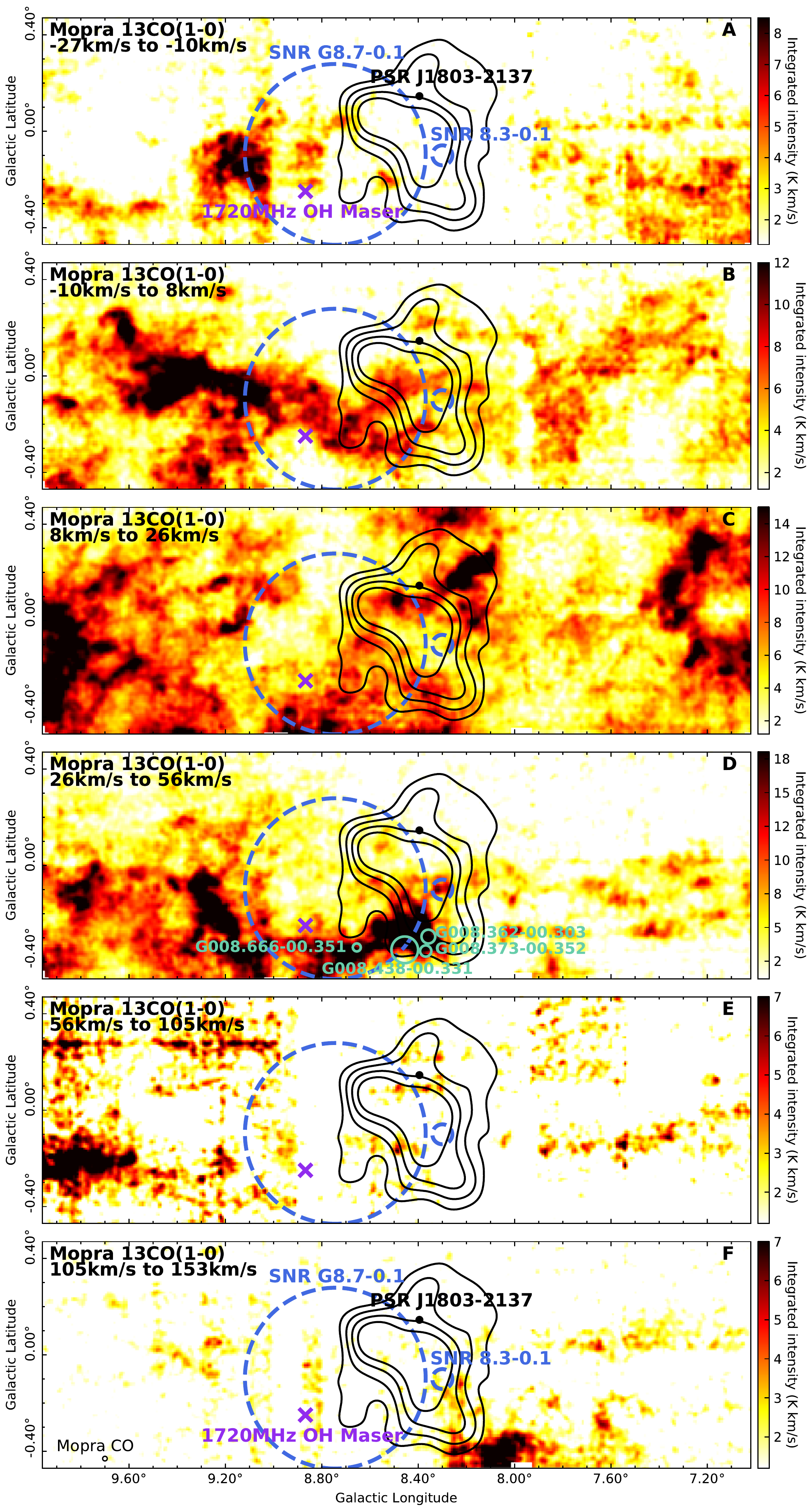}
\caption{Mosaic of Mopra $^{13}$CO integrated intensity maps (K\,\kms) towards \hessj. The two dashed blue circles indicate \snrg and \snr. The \OH is indicated by the purple cross and \psr is indicated by the black dot. The TeV $\gamma$-ray emission for 5-10$\sigma$ is shown by the solid black contours. The aqua circles in component D indicate H\textsc{ii} regions.}
\label{fig:13CO_mop}
\end{center}
\end{figure*}

\begin{figure*}
\begin{center}
\includegraphics[width=1.4\columnwidth]{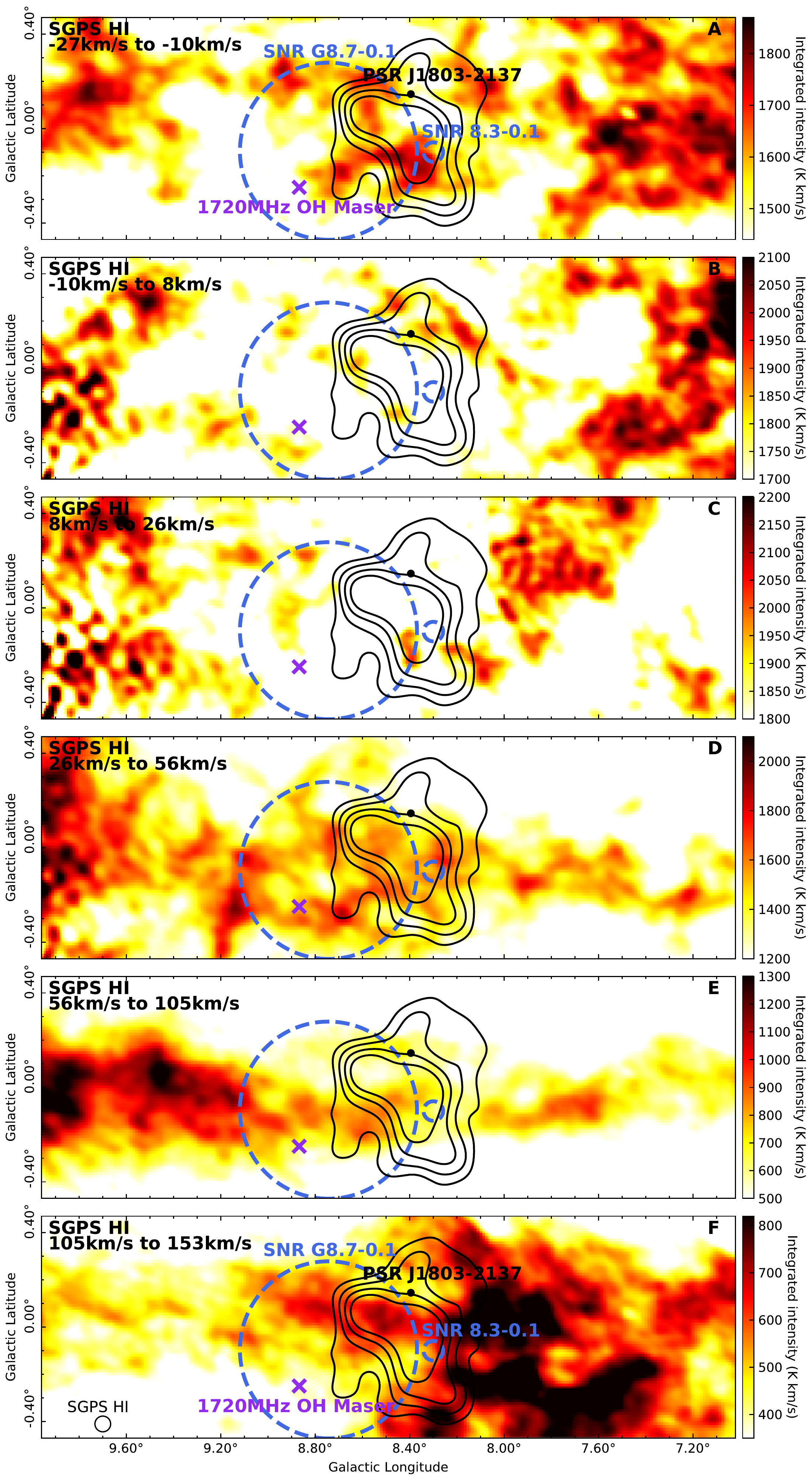}
\caption{Mosaic of SGPS HI integrated intensity maps (K\,\kms) towards \hessj. The two dashed blue circles indicate \snrg and \snr. The \OH is indicated by the purple cross and \psr is indicated by the black dot.  The TeV $\gamma$-ray emission for 5-10$\sigma$ is shown by the solid black contours.}
\label{fig:HI_sgps}
\end{center}
\end{figure*}

\begin{figure*}
\begin{center}
\includegraphics[width=1.4\columnwidth]{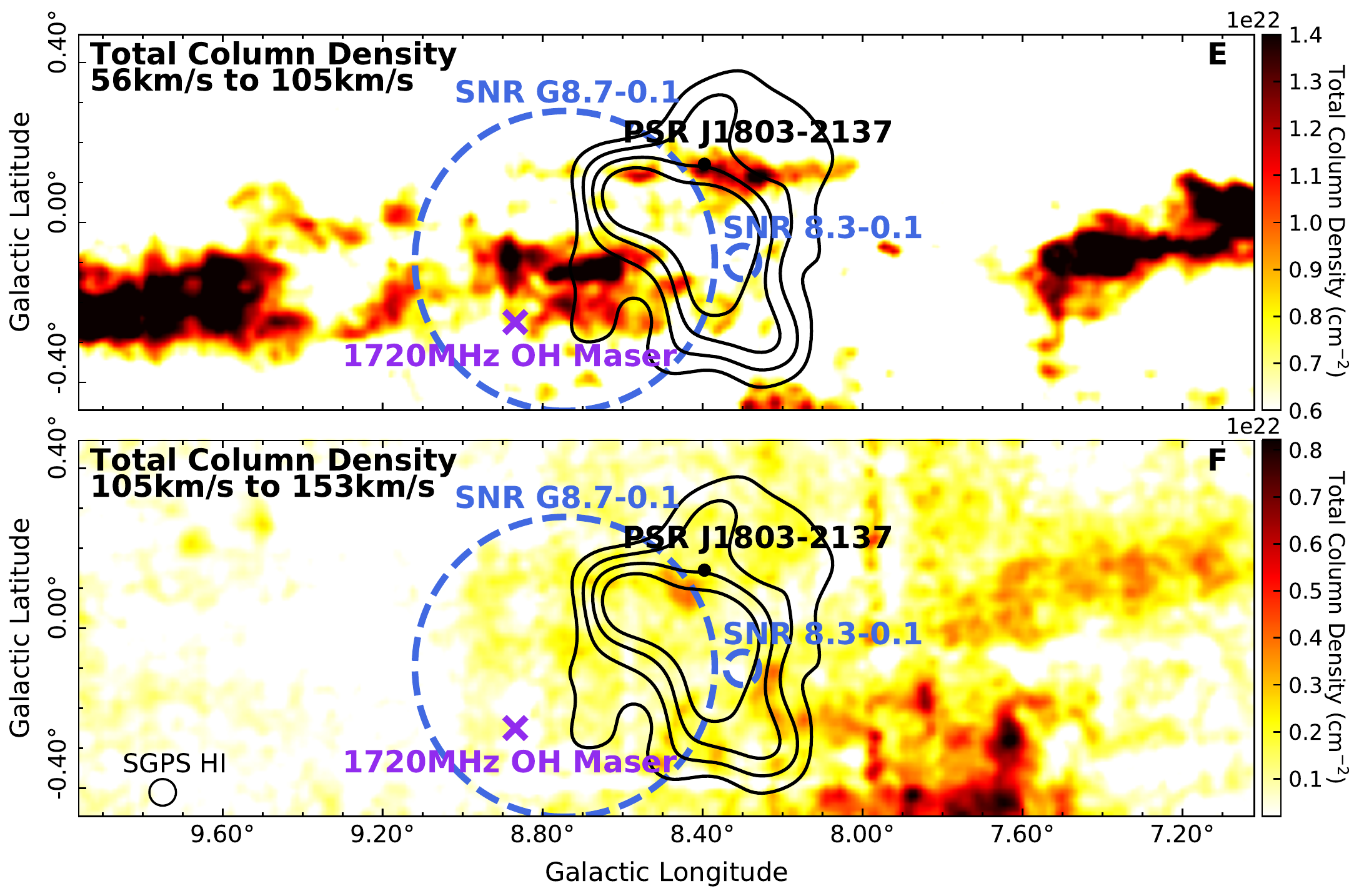}
\caption{Total column density maps, $2N_{\rm{H_2}}+N_{\rm{HI}}$, ($\rm{cm}^{-2}$) towards \hessj, for gas components E and F. The two dashed blue circles indicate \snrg and \snr. The \OH is indicated by the purple cross and \psr is indicated by the black dot. The TeV $\gamma$-ray emission for 5-10$\sigma$ is shown by the solid black contours.}
\label{fig:total_E+F}
\end{center}
\end{figure*}

\begin{figure*}
\begin{center}
\includegraphics[width=1.4\columnwidth]{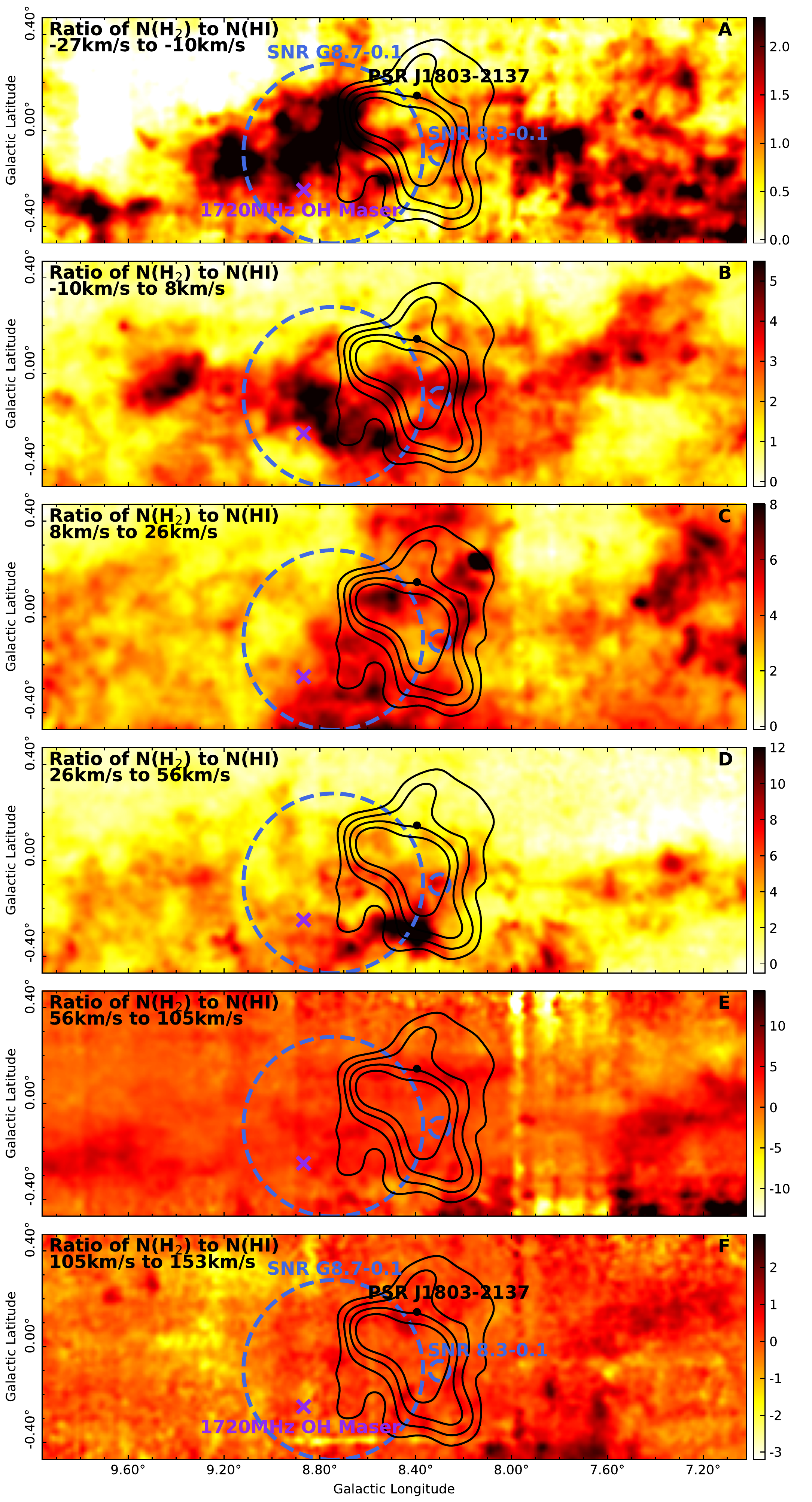}
\caption{Ratio of molecular hydrogen ($N_{\rm{H_2}}$) and atomic hydrogen ($N_{\rm{HI}}$) column densities towards \hessj, for gas components A-E. The two dashed blue circles indicate \snrg and \snr. The \OH is indicated by the purple cross and \psr is indicated by the black dot. The TeV $\gamma$-ray emission for 5-10$\sigma$ is shown by the solid black contours.}
\label{fig:ratio}
\end{center}
\end{figure*}

The ISM is made up of both atomic and molecular gas, primary HI and $^{12}$CO emission respectively. However, there are regions in which these gas tracers become `invisible', due to a lack of emission. It has been shown that there is a component of gas which has not been detected, commonly known as `dark' gas \citep{Dark_Li_2018}.

In addition to the common neutral gas tracers (HI and $^{12}$CO), a component of ionised gas is present in interstellar clouds. For cases in which clouds are optically thick, the dust opacity maps from the Planck collaboration \citep{Planck_collab_2016} can be used to estimate a hydrogen column density \citep{Eq_Planck_2011}. The column density derived via this method contains no distance information as the dust opacity map has been summed over the line-of-sight. The Planck hydrogen column density is therefore an upper limit. 

\begin{figure*}
\begin{center}
\includegraphics[width=1.8\columnwidth]{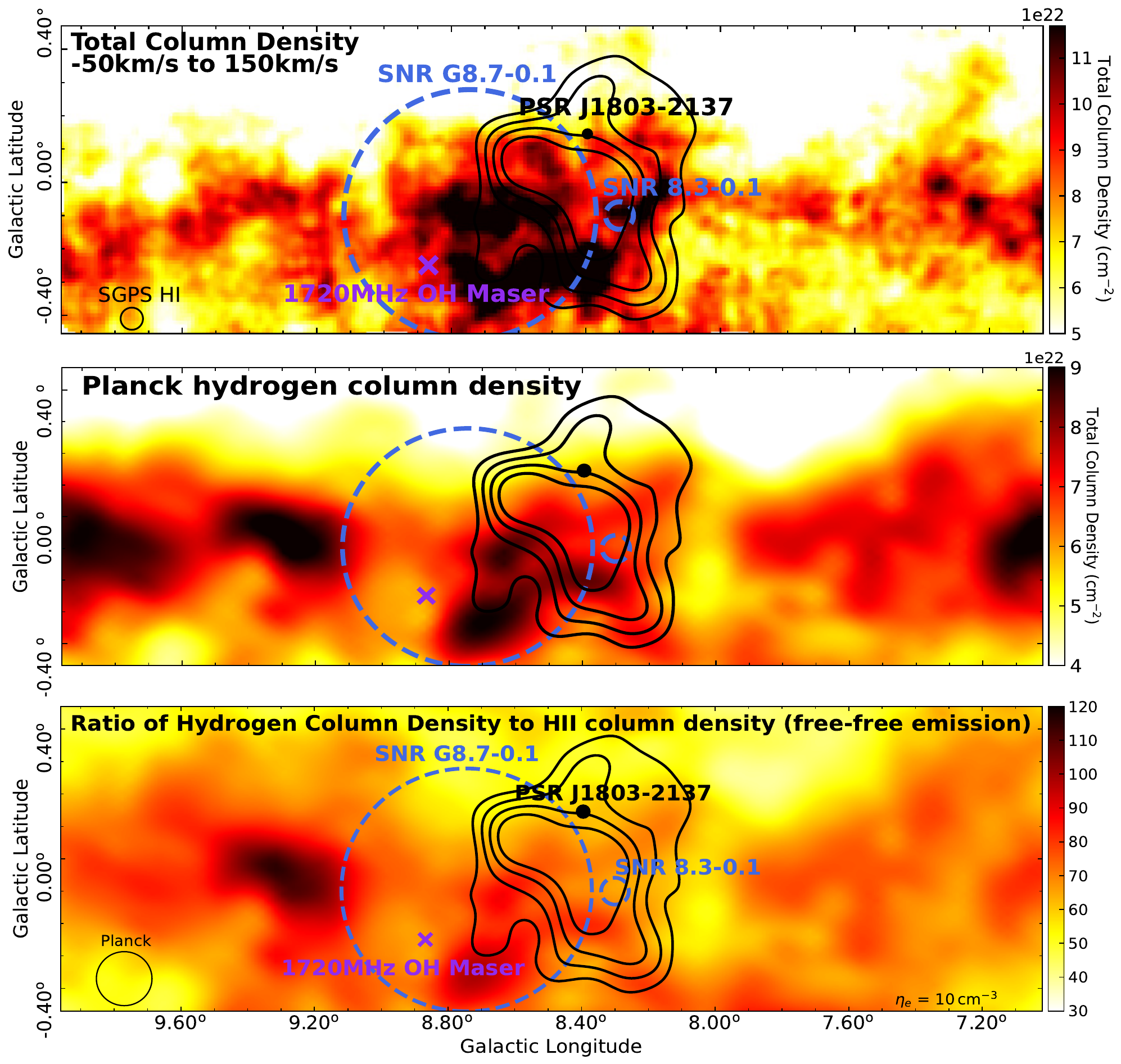}
\caption{\textit{Top:} Total column density map ($\rm{cm}^{-2}$) from the SGPS HI and Mopra $^{12}$CO emission, along the entire line-of-sight (\vlsr$=50$\ to\ $150$\,\kms) of \hessj. \textit{Middle}: Planck hydrogen column density. \textit{Bottom}: Ratio of hydrogen column density as derived from Planck dust opacity and H\textsc{ii} column density from free-free emission. \textit{All}: The two dashed blue circles indicate \snrg and \snr. The \OH is indicated by the purple cross and \psr is indicated by the black dot. The TeV $\gamma$-ray emission for 5-10$\sigma$ is shown by the solid black contours.}
\label{fig:planck}
\end{center}
\end{figure*}

To determine the Planck H\textsc{ii} column density the free-free emission map was required \citep{Planck_collab_2016}. To convert the emission map into a free-free intensity map, the conversion factor $I_{\nu} = 46.04$\,Jy\,sr$^{-1}$ at 353\,GHz \citep[from][]{Table_Finkbeiner_2003} was applied.  Equation~5 from \cite{HII_Sodroski_1997} is then used to derive the H\textsc{ii} column density. Here we use an effective electron density ($n_{\rm{eff}}$) of $10\,\rm{cm}^{-3}$ as a lower limit. Similarly to the Planck hydrogen column density, the H\textsc{ii} column density is integrated along the whole line-of-sight. The bottom panel of Figure~\ref{fig:planck} shows the ratio between column density derived from the dust opacity map  and column density from free-free emission (H\textsc{ii} column density).

The total hydrogen column density traced by the HI and $^{12}$CO emission, is taken along the entire line-of-sight (\vlsr$=50$\ to\ $150$\,\kms) to allow for comparison between it and the Planck data, which has no distance information. 

The total hydrogen column density (top panel of Figure~\ref{fig:planck}) has morphological similarities to the total column density as derived from the dust opacity map, as demonstrated in the middle panel of Figure~\ref{fig:planck}. In particular, we note the dense region of gas to the Galactic-South of the TeV source present in both column density maps. They are also on the same order of magnitude, hence to compare the H\textsc{ii} column density with the total hydrogen column density it is acceptable to use the dust opacity column density. This ratio is presented in the bottom panel of Figure~\ref{fig:planck}. The ratio values indicate that the total neutral column density is dominating over the component of ionised gas. For this purpose, the total column density used through-out this paper does not take the ionised gas into account.

\subsection{Dense Gas Tracer Mosaics}
\label{subsec:mos_dense}
Table~\ref{tab:7mm} shows the molecular lines that were observed by the 7\,mm observing set-up for the Mopra Spectrometer.

\begin{table}[!h]
\centering
\caption{Molecular lines with each of their rest line frequencies from the 7\,mm receiver of the Mopra telescope.}
\begin{tabular}{cc} 
\hline 
Molecular Line        & Line Rest Frequency (GHz) \\ 
\hline
$^{30}$SiO(1-0, v=0) & 42.373365           \\
SiO(1-0, v=3)        & 42.519373           \\
SiO(1-0, v=2)        & 42.820582           \\
SiO(1-0, v=1)        & 43.122079           \\
SiO(1-0, v=0)        & 43.423864           \\
CH$_3$OH(I)          & 44.069476           \\
HC$_7$N(40-39)       & 45.119064           \\
HC$_5$N(17-16)       & 45.264750           \\
HC$_3$N(5-4, F=5-5)  & 45.488839           \\
$^{13}$CS(1-0)       & 46.247580           \\
HC$_5$N(16-15)       & 47.927275           \\
C$^{34}$S(1-0)       & 48.206946           \\
OCS(4-3)             & 48.651604           \\
CS(1-0)              & 48.990957           \\
\hline
\end{tabular}
\label{tab:7mm}
\end{table}

\begin{figure*}
\begin{center}
\includegraphics[width=1.9\columnwidth]{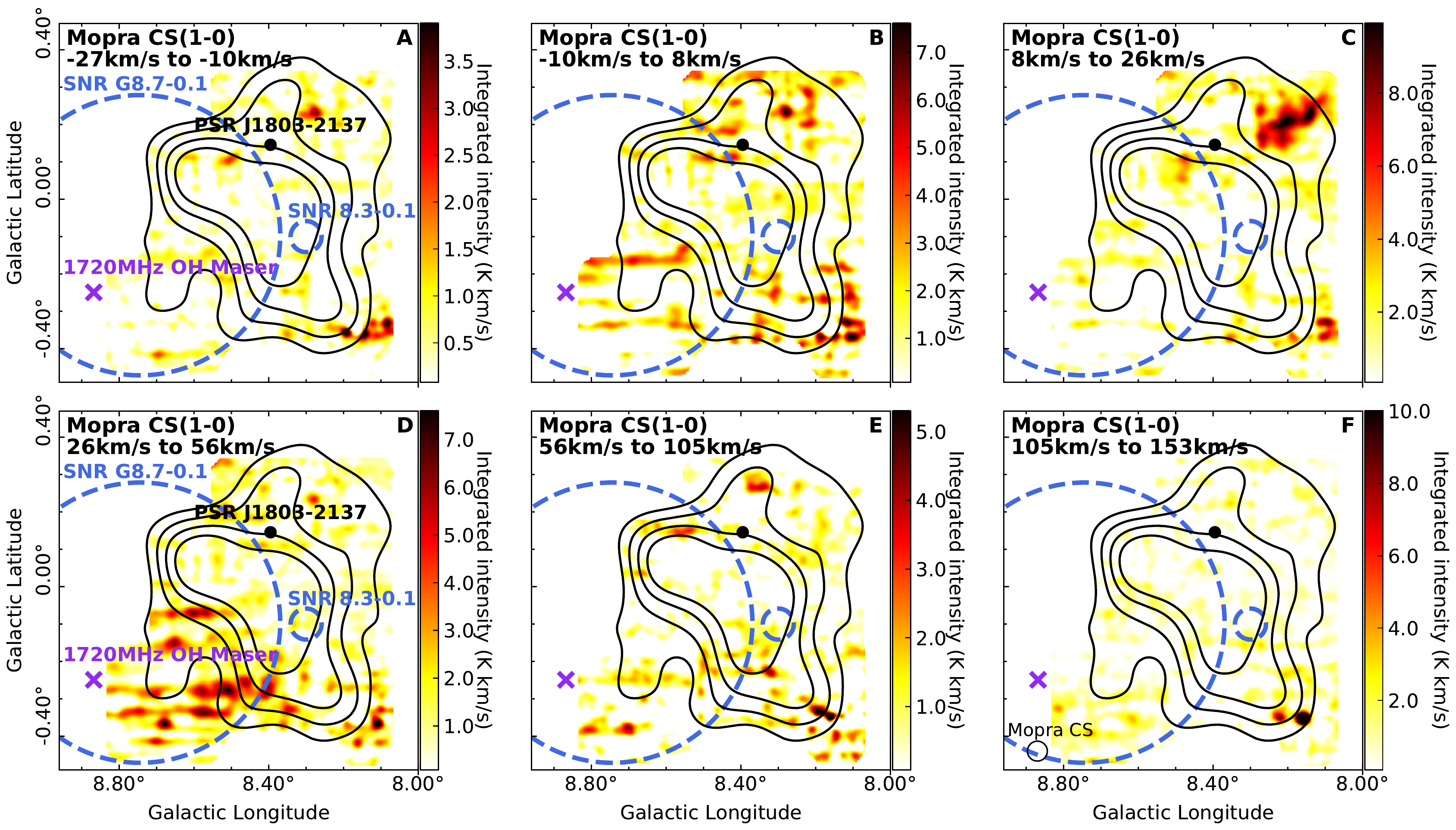}
\caption{CS(1-0) integrated intensity maps ($\rm{K\,km\,s}^{-1}$) towards \hessj. For components A through F the T$_{\rm{rms}}$ is 2.5\,K, 2.7\,K, 2.8\,K, 3.5\,K, 4.3\,K and 4.2\,K respectively.
The two dashed blue circles indicate \snrg and \snr. The \OH is indicated by the purple cross and \psr is indicated by the black dot. The TeV $\gamma$-ray emission for 5-10$\sigma$ is shown by the solid black contours.}
\label{fig:CS}
\end{center}
\end{figure*}

\begin{figure*}
\begin{center}
\includegraphics[width=1.9\columnwidth]{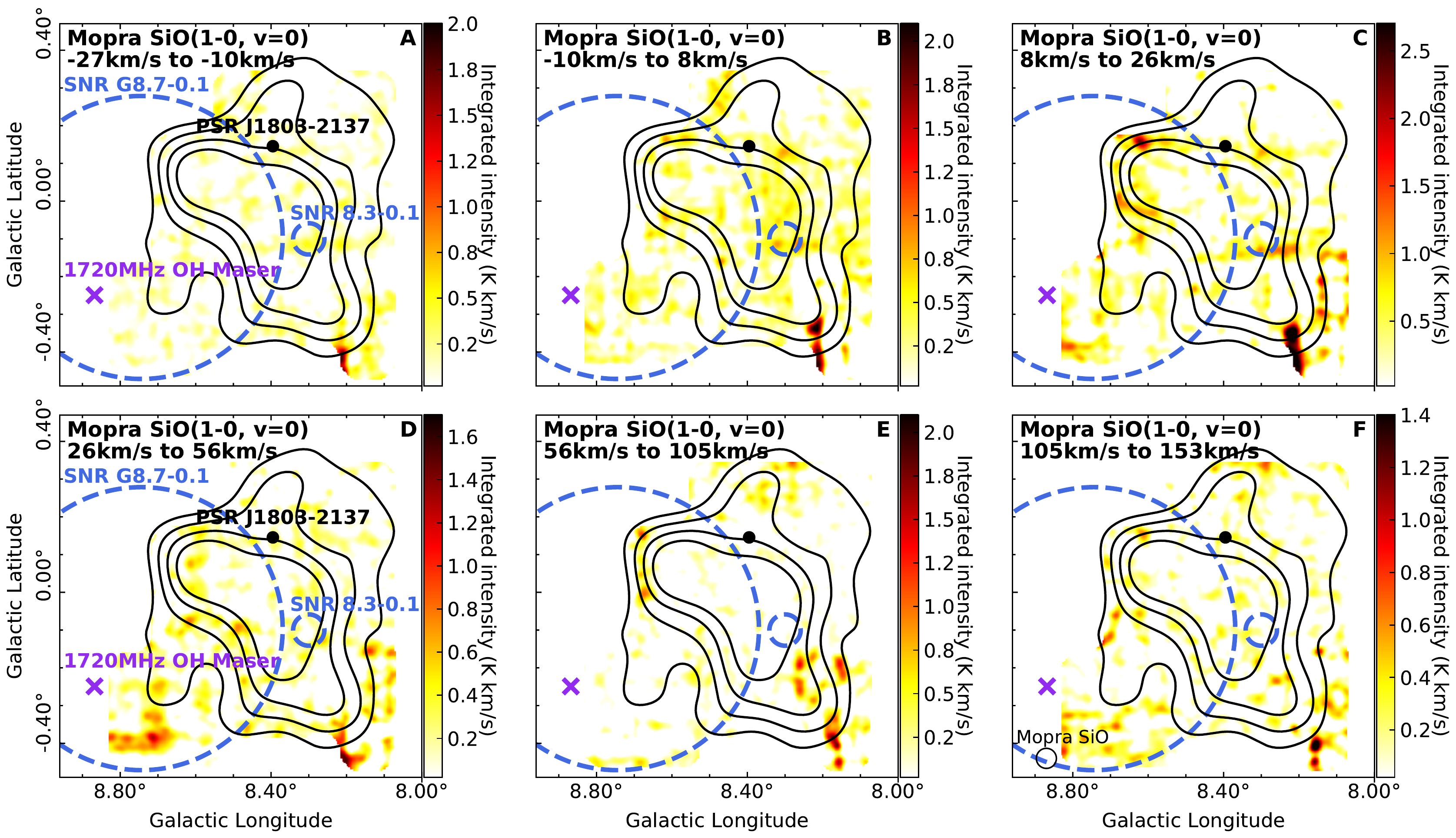}
\caption{SiO(1-0, v=0) integrated intensity maps ($\rm{K\,km\,s}^{-1}$) towards \hessj. For components A through F the T$_{\rm{rms}}$ is 1.3\,K, 1.3\,K, 1.4\,K, 1.7\,K, 2.2\,K and 2.2\,K respectively.
The two dashed blue circles indicate \snrg and \snr. The \OH is indicated by the purple cross and \psr is indicated by the black dot. The TeV $\gamma$-ray emission for 5-10$\sigma$ is shown by the solid black contours.}
\label{fig:SiO}
\end{center}
\end{figure*}

\begin{figure*}
\begin{center}
\includegraphics[width=1.9\columnwidth]{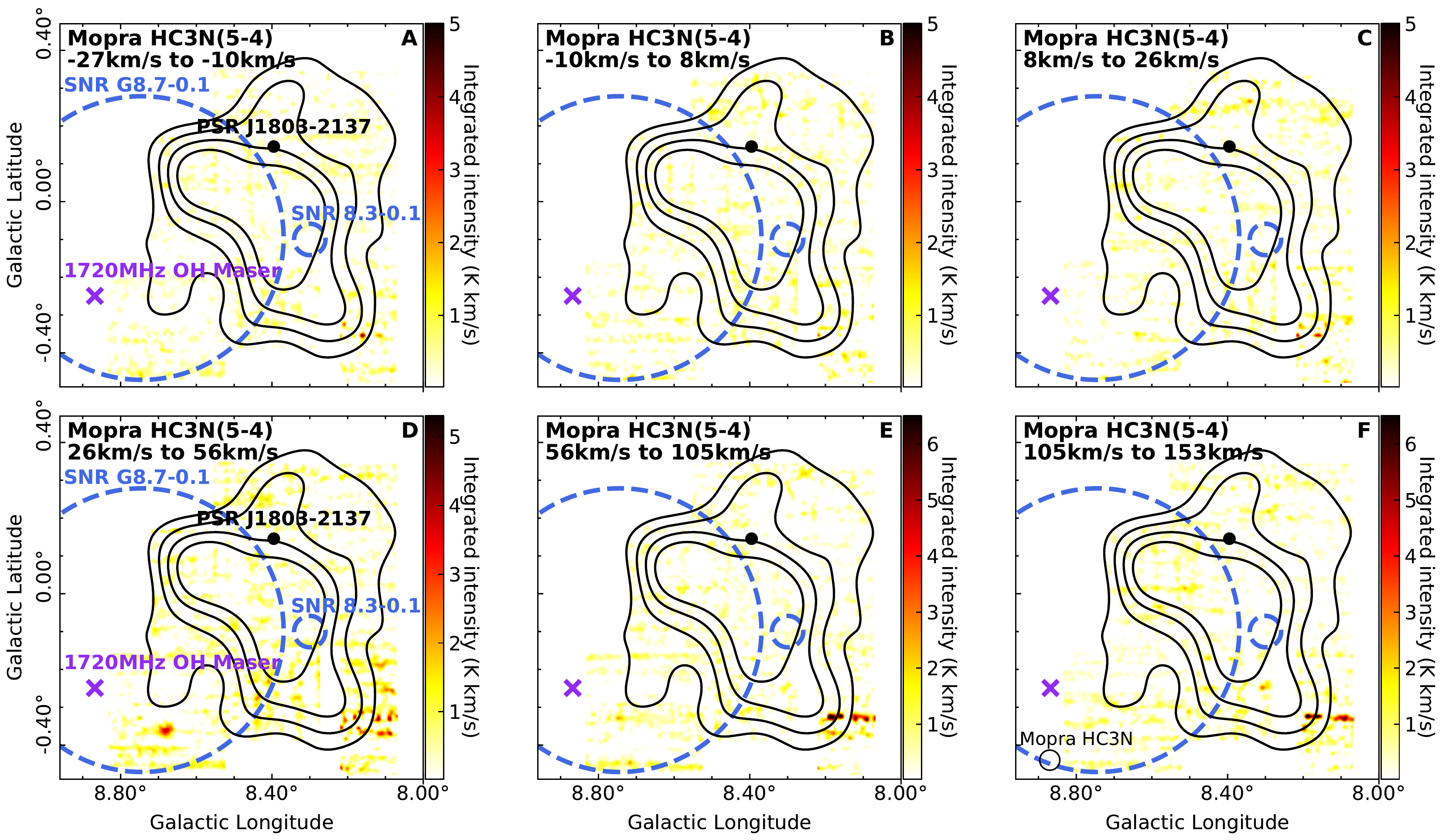}
\caption{HC$_3$N(5-4) integrated intensity maps ($\rm{K\,km\,s}^{-1}$) towards \hessj. For components A through F the T$_{\rm{rms}}$ is 3.6\,K, 3.8\,K, 3.8\,K, 4.9\,K, 6.1\,K and 6.1\,K respectively.
The two dashed blue circles indicate \snrg and \snr. The \OH is indicated by the purple cross and \psr is indicated by the black dot. The TeV $\gamma$-ray emission for 5-10$\sigma$ is shown by the solid black contours.}
\label{fig:HC3N}
\end{center}
\end{figure*}

\begin{figure*}
\begin{center}
\includegraphics[width=1.9\columnwidth]{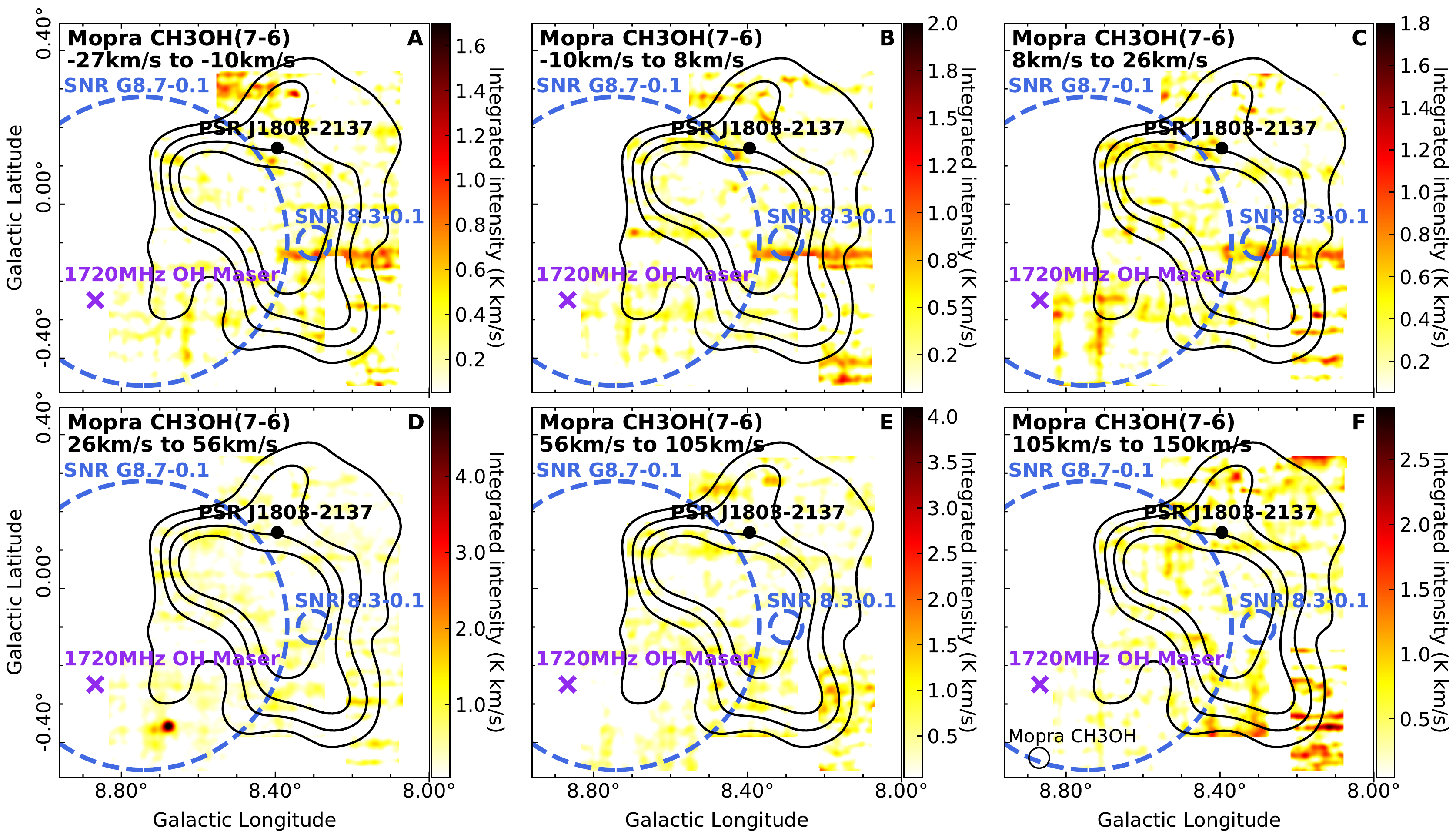}
\caption{CH$_3$OH integrated intensity maps ($\rm{K\,km\,s}^{-1}$, uncleaned) towards \hessj. For components A through F the T$_{\rm{rms}}$ is 0.8\,K, 0.9\,K, 0.9\,K, 1.1\,K, 1.4\,K and 1.4\,K respectively.
The two dashed blue circles indicate \snrg and \snr. The \OH is indicated by the purple cross and \psr is indicated by the black dot. The TeV $\gamma$-ray emission for 5-10$\sigma$ is shown by the solid black contours.}
\label{fig:CH3OH}
\end{center}
\end{figure*}

\begin{figure*}
\begin{center}
\includegraphics[width=1.4\columnwidth]{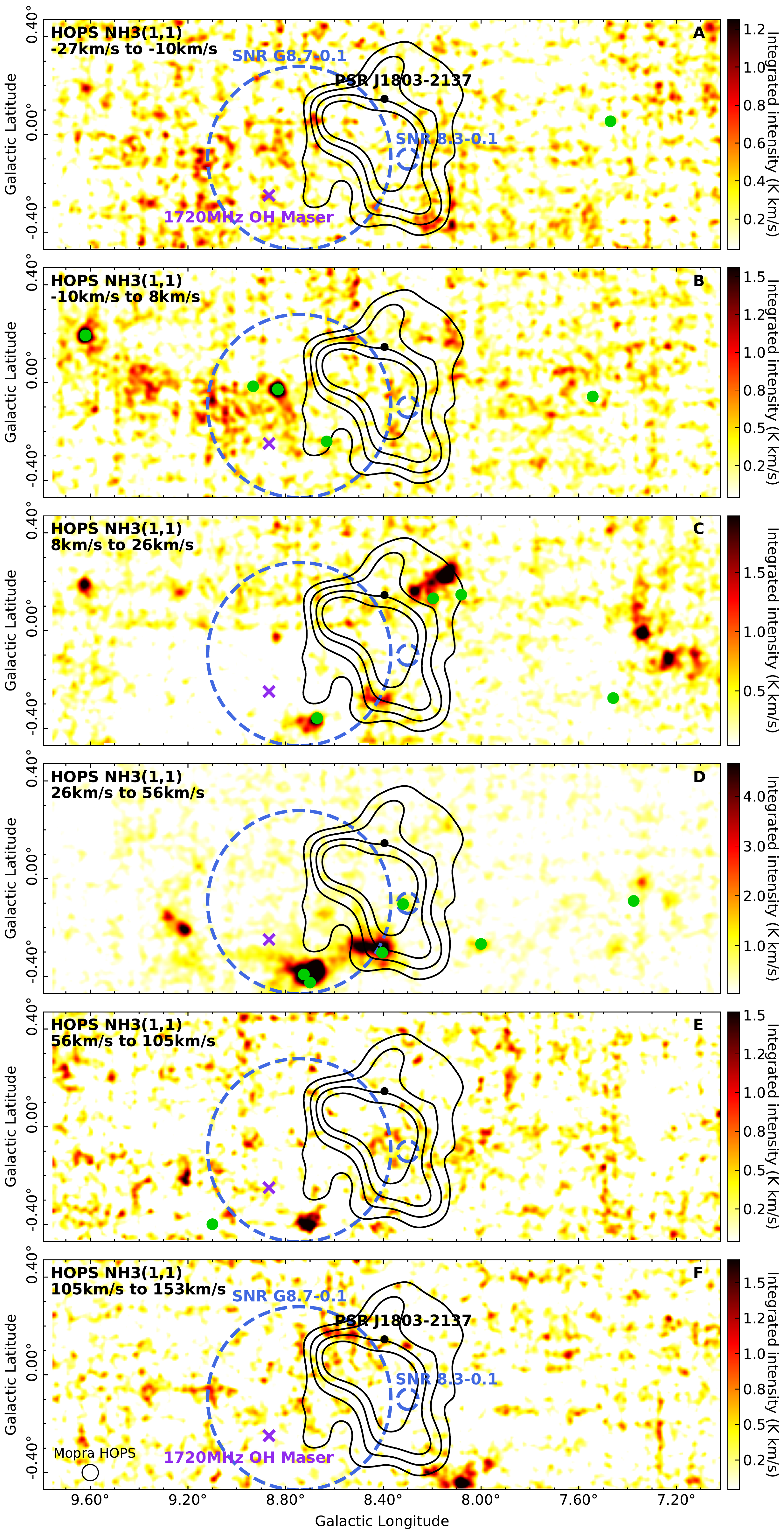}
\caption{NH$_3$(1,1) integrated intensity maps ($\rm{K\,km\,s}^{-1}$) towards \hessj using HOPS data. The two dashed blue circles indicate \snrg and \snr. The \OH is indicated by the purple cross and \psr is indicated by the black dot. The TeV $\gamma$-ray emission for 5-10$\sigma$ is shown by the solid black contours. H$_2$O maser positions are shown by the green dots.}
\label{fig:NH3}
\end{center}
\end{figure*}

The integrated intensity maps for the various dense gas tracers towards \hessj are presented in Figures~\ref{fig:CS},~\ref{fig:SiO},~\ref{fig:HC3N},~\ref{fig:CH3OH}~and~\ref{fig:NH3}.

SiO(1-0, v=0) emission has been detected towards \hessj, however it is quite weak. There are a few dense features in components B, C and D, however these show features which have already been seen in the other dense gas tracers (see Section~\ref{subsec:ISM_comps}).

%

In Figure~\ref{fig:NH3} we have included the known H$_2$O maser positions from \cite{NH3_Walsh_2011}, at their given velocities.

\FloatBarrier
\section{CR spectra model}
\label{subsec:A_A_plots}
Equation~\ref{eqn:CR_flux_k} can be adjusted to calculate the CR enhancement factor from the GeV $\gamma$-rays from \fges, as shown by Equation~\ref{eqn:CR_flux_k2}. An integral power law spectrum of $E^{-1.75}$ is assumed, following \cite{VHE_Aharonian_1991} for GeV energies:

\begin{equation}
F(\geq E_{\gamma})=1.45 \times 10^{-13} E_{\rm{TeV}}^{-1.75} \left( \dfrac{M_5}{d_{\rm{kpc}}^2} \right) k_{\mathrm{CR}} \quad \rm{cm}^{-2}\,\rm{s}^{-1}
\label{eqn:CR_flux_k2}
\end{equation}

The photon flux for $\gamma$-rays from \fges is $F(\ge10\,\rm GeV)=1.56\times 10^{-9}\,\rm{cm}^{-2}\,\rm{s}^{-1}$ \citep{Fermi_Ackermann_2017}. This leads to a CR enhancement factor, $k_{\rm CR}$, of $\mathord{\sim}9$ times that of the Earth-like CR density for \snrg (component D) at GeV energies. 

\begin{figure*}
\includegraphics[width=2\columnwidth]{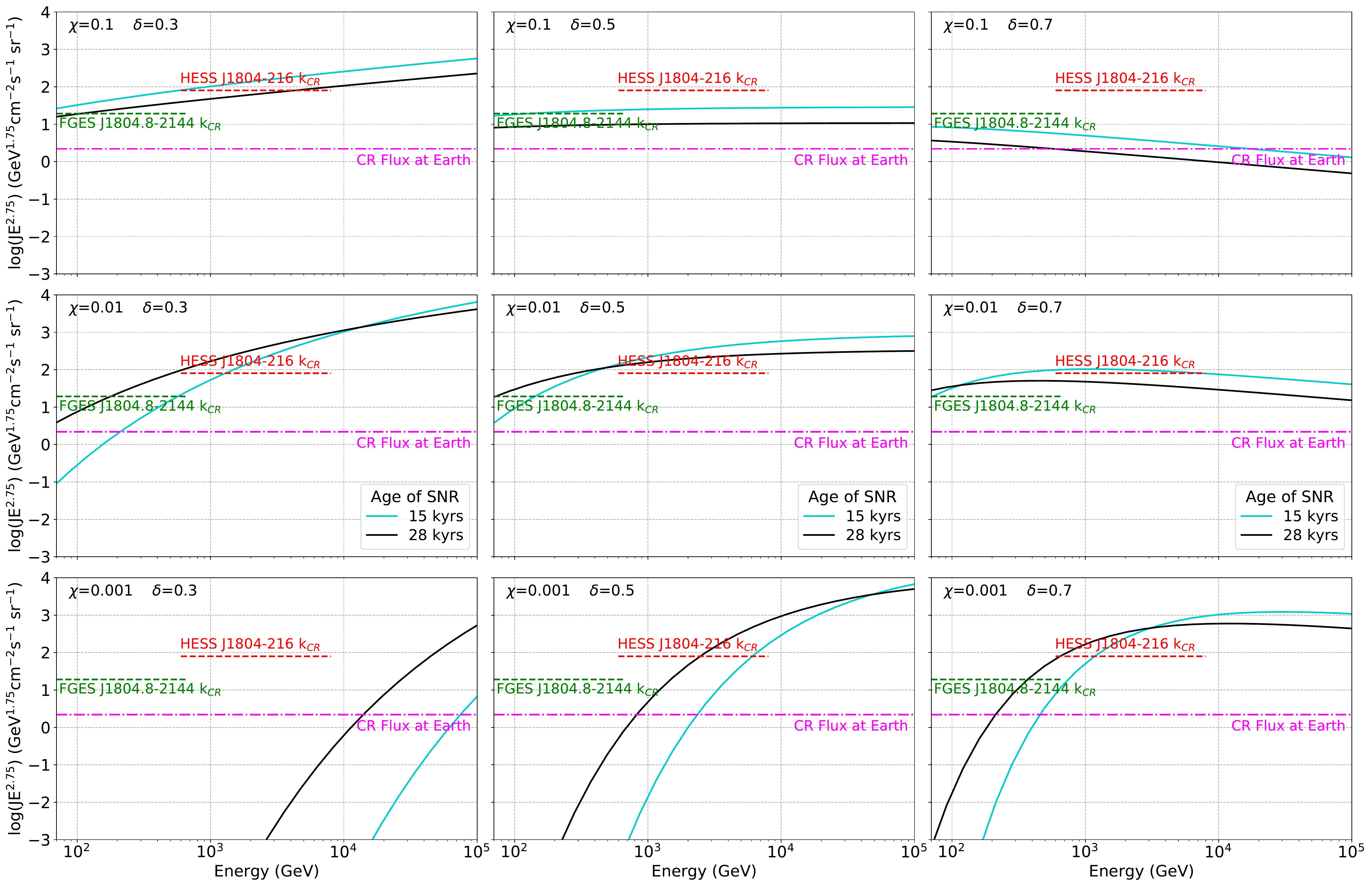}
\caption{Modelled energy spectra of CR protons (Equation~\ref{eqn:flux_dif}) escaping from the potential impulsive accelerator \snrg, with a total energy of $2\times10^{48}$\,erg in CRs. Various values of diffusion suppression factor, $\chi$, and index of the diffusion coefficient, $\delta$, are shown here. A power-law spectrum with a spectral index of $\alpha=2$ is assumed. The number density is taken to be $n=160$\,cm$^{-3}$. The distance from the accelerator to the cloud is $R\mathord{\sim}12$\,pc and age of the source are taken to be 15\,kyr and 28\,kyr for the cyan and black curves, respectively. The magenta dashed line represents the CR flux observed at Earth. The red represents the calculated CR enhancement factor for \hessj ($k_{\rm CR}\approx37$). The green represents the calculated CR enhancement factor for \fges ($k_{\rm CR}\approx9$).}
\label{fig:aa_vary_delta}
\end{figure*}

Using Equation~\ref{eqn:flux_dif} the energy spectrum of CR protons is obtained for a range of diffusion suppression factors, $\chi$'s, and indices of the diffusion coefficient, $\delta$'s, to test the validity of each value, as shown in Figure~\ref{fig:aa_vary_delta}.

Values of $\delta=0.5\ \& \ 0.7$ for $\chi=0.01$ are the most plausible for the hadronic scenario for \snrg (see Section~\ref{subsec:hadronic}).

\begin{figure*}
\includegraphics[width=2\columnwidth]{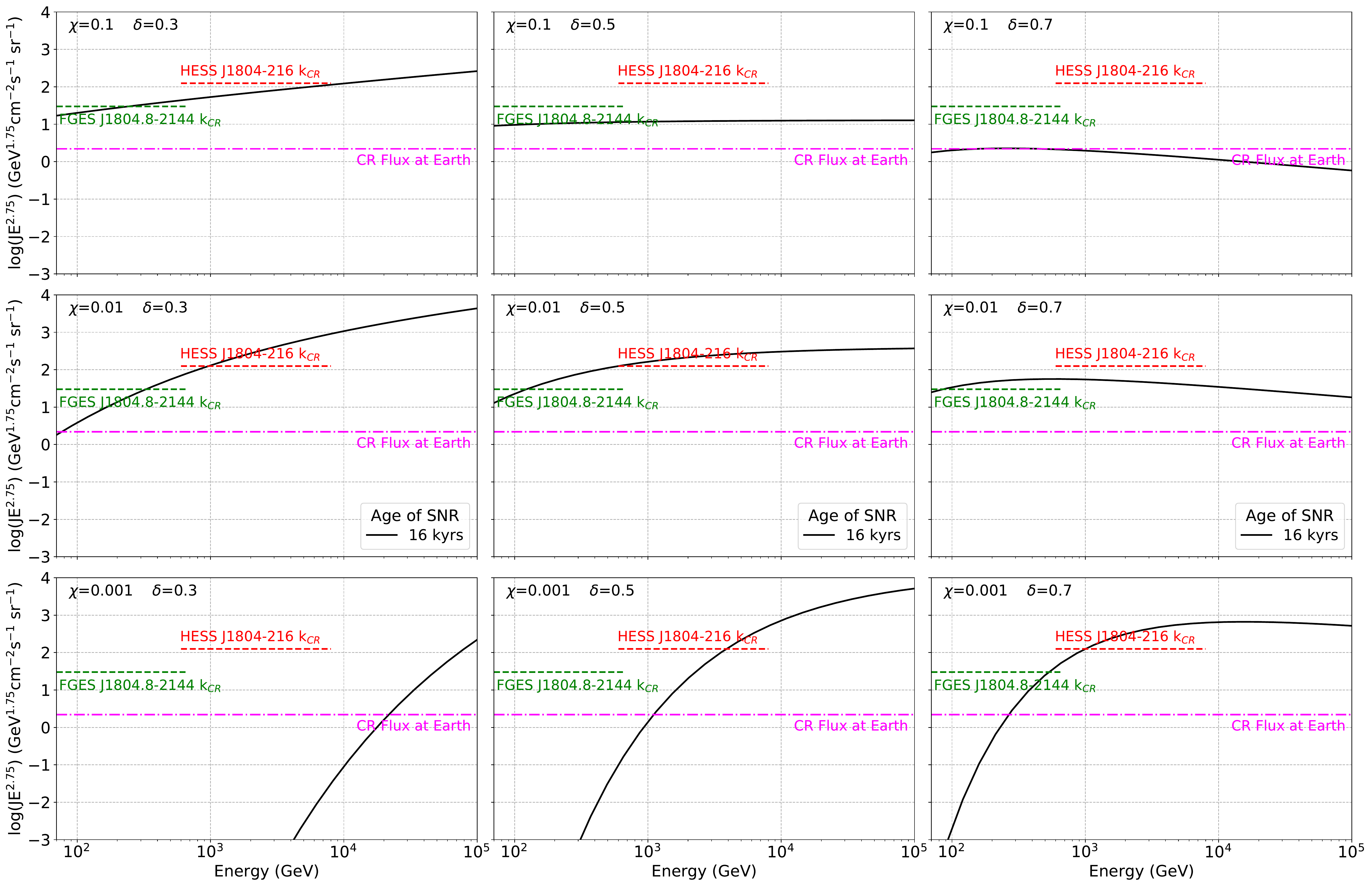}
\caption{Modelled energy spectra of CR protons (Equation~\ref{eqn:flux_dif}) escaping from the potential impulsive accelerator (progenitor SNR from \psr), with a total energy of $10^{48}$\,erg in CRs. Various values of diffusion suppression factor, $\chi$, and index of the diffusion coefficient, $\delta$, are shown here. A power-law spectrum with a spectral index of $\alpha=2$ is assumed. The number density is taken to be $n=325$\,cm$^{-3}$. The distance from the accelerator to the cloud is $R\mathord{\sim}10$\,pc and age of the source is taken to be 16\,kyr for the black curves. The magenta dashed line represents the CR flux observed at Earth. The red represents the calculated CR enhancement factor for \hessj ($k_{\rm CR}\approx57$). The green represents the calculated CR enhancement factor for \fges ($k_{\rm CR}\approx14$).}
\label{fig:aa_psr}
\end{figure*}

Figure~\ref{fig:aa_psr} shows the energy spectrum of CR protons escaping from the progenitor SNR of \psr. The total energy budget of CRs in this scenario is taken to be $10^{48}$\,erg which is consistent with $W_{p,\rm{TeV}}$ from Equation~\ref{eqn:tot_E_budget} (see also Table~\ref{tab:tot_CR_budget}) using component C.
A CR enhancement factor, $k_{\rm CR}$, of $\mathord{\sim}57$ is obtained for TeV energies (using Equation~\ref{eqn:CR_flux_k}) and $\mathord{\sim}14$ for GeV energies (using Equation~\ref{eqn:CR_flux_k2}).

\end{document}